\documentclass[aps,prd,unsortedaddress,reprint,nofootinbib]{revtex4-1}
\pdfoutput=1 
\usepackage{graphicx}
\usepackage[dvipsnames]{xcolor}
\usepackage{color}
\definecolor{darkblue}{rgb}{0.0, 0.0, 0.55}
\definecolor{cite}{rgb}{0.0, 0.34, 0.25}

\usepackage{rotating}
\usepackage{bigints}
\usepackage{lineno}

\newcommand{\beq}{\begin{equation}}
\newcommand{\eeq}{\end{equation}}
\newcommand{\bea}{\begin{eqnarray}}
\newcommand{\eea}{\end{eqnarray}}
\newcommand \hmu {\hat{\mu}}

\newcommand{\fourback}{\! \! \! \!}
\newcommand{\calZ}{{\cal Z}}
\newcommand{\MeV}{\, {\rm MeV}}
\newcommand{\GeV}{\, {\rm GeV}}

\begin{document}
\title{Off-diagonal correlators of conserved charges from lattice QCD, and  how to relate them to experiment}

\author{R. Bellwied}
\affiliation{Department of Physics, University of Houston, Houston, TX 77204, USA}

\author{S. Bors\'anyi}
\affiliation{University of Wuppertal, Department of Physics, Wuppertal D-42119, Germany}

\author{Z. Fodor}
\affiliation{University of Wuppertal, Department of Physics, Wuppertal D-42119, Germany}
\affiliation{E{\"o}tv{\"o}s University, Budapest 1117, Hungary}
\affiliation{J{\"u}lich Supercomputing Centre, J{\"u}lich D-52425, Germany}
\affiliation{UCSD, Physics Department, San Diego, CA 92093, USA}

\author{J. N. Guenther}
\affiliation{University of Wuppertal, Department of Physics, Wuppertal D-42119, Germany}
\affiliation{University of Regensburg, Department of Physics, Regensburg D-93053, Germany}

\author{J. Noronha-Hostler}
\affiliation{Department of Physics, 
University of Illinois at Urbana-Champaign, Urbana, IL 61801, USA}

\author{P. Parotto}
\email[Corresponding author: ]{\url{parotto@uni-wuppertal.de}}
\affiliation{Department of Physics, University of Houston, Houston, TX 77204, USA}
\affiliation{University of Wuppertal, Department of Physics, Wuppertal D-42119, Germany}

\author{A. P\'asztor}
\affiliation{E{\"o}tv{\"o}s University, Budapest 1117, Hungary}

\author{C. Ratti}

\author{J. M. Stafford}
\affiliation{Department of Physics, University of Houston, Houston, TX 77204, USA}

\date{\today}

\begin{abstract}
Like fluctuations, non-diagonal correlators of conserved charges provide a tool
for the study of chemical freeze-out in heavy ion collisions. They can be
calculated in thermal equilibrium using lattice simulations, and be connected to moments of 
event-by-event net-particle multiplicity distributions. 
We calculate them from
continuum extrapolated lattice simulations at $\mu_B=0$, and present a
finite-$\mu_B$ extrapolation, comparing two different methods. 
In order to relate the grand canonical observables to the
experimentally available net-particle fluctuations and correlations, we perform
a Hadron Resonance Gas (HRG) model analysis, which allows us to completely
break down the contributions from different hadrons. We then construct suitable
hadronic proxies for fluctuations ratios, and study their behavior at finite
chemical potentials. We also study the effect of introducing acceptance cuts,
and argue that the small dependence of certain ratios on the latter allows for
a direct comparison with lattice QCD results, provided that the same cuts are
applied to all hadronic species. Finally, we perform a comparison for the
constructed quantities for experimentally available measurements from the STAR
Collaboration. Thus, we estimate the chemical freeze-out temperature to 165~MeV
using a strangeness-related proxy. This is a rather high temperature
for the use of the Hadron Resonance Gas, thus, further lattice
studies are necessary to provide first principle results at intermediate $\mu_B$.
\end{abstract}
\pacs{}

\maketitle

\section{Introduction}

The study of the phase diagram of Quantum Chromodynamics (QCD) has been the object of intense effort from both theory and experiment in the last decades. Relativistic heavy ion collision experiments both at the Relativistic Heavy Ion Collider (RHIC) and the Large Hadron Collider (LHC) have been able to create the Quark Gluon Plasma (QGP) in the laboratory, and explore the low-to-moderate baryon density region of the QCD phase diagram.

At low baryon density, the transition from a hadron gas to a deconfined QGP was
shown by lattice QCD calculations to be a broad crossover \cite{Aoki:2006we} at
$T \simeq 155 \MeV$
\cite{Aoki:2006we,Aoki:2009sc,Borsanyi:2010bp,Bazavov:2011nk}. At large baryon
densities, the nature of the phase transition is expected to change into first
order, thus implying the presence of a critical end point. A strong experimental
effort is currently in place through the second Beam Energy Scan (BES-II)
program at RHIC in 2019-2021, with the goal of discovering such a critical
point.

The structure of the QCD phase diagram cannot currently be theoretically
calculated from first principles, as lattice calculations are hindered by the
sign problem at finite density. Several methods have been utilized in order to
expand the reach of lattice QCD to finite density, like full reweighting
\cite{Fodor:2001au}, Taylor expansion of the observables around $\mu_B=0$
\cite{Allton:2002zi, Allton:2005gk, Gavai:2008zr, Basak:2009uv,
Kaczmarek:2011zz}, or their analytical continuation from imaginary chemical
potential
\cite{Fodor:2001au,deForcrand:2002hgr,DElia:2002tig,Fodor:2001pe,Fodor:2004nz}. 

We remark here, that there are alternative approaches to lattice QCD for
the thermodynamical description. Specific truncations of the
Dyson-Schwinger equations allow the calculation of the crossover line and also
to extract baryonic fluctuations \cite{Fischer:2018sdj,Isserstedt:2019pgx}. Another theoretical result on the baryon-strangeness correlator has been
calculated using functional methods from the Polyakov-loop-extended quark meson
model in \cite{Fu:2018qsk,Fu:2018swz}.

The confined, low-temperature regime of the theory is well described by the
Hadron Resonance Gas (HRG) model, which is able to reproduce the vast majority
of lattice QCD results in this regime
\cite{Dashen:1969ep,Venugopalan:1992hy,Karsch:2003vd,Karsch:2003zq,Tawfik:2004sw,Ratti:2010kj}.
Moreover, the HRG model has been extremely successful in reproducing
experimental results for particle yields over several orders of magnitude
\cite{Cleymans:1998yb, Andronic:2008gu, Manninen:2008mg, Abelev:2009bw,
Aggarwal:2010pj}. These are usually referred to as thermal fits, since the goal
of the procedure is the determination of the temperature and chemical potential
at which the particle yields are frozen. This moment in the evolution of a
heavy ion collision is called chemical freeze-out, and takes place when
inelastic collisions within the hot hadronic medium cease. The underlying
assumption is that the system produced in heavy ion collisions eventually
reaches thermal equilibrium
\cite{Rapp:2000gy,BraunMunzinger:2003zz,NoronhaHostler:2007jf}, and therefore a
comparison between thermal models and experiment is possible
\cite{Ratti:2018ksb} . 

Although the net number of individual particles may change after the 
chemical freeze-out through resonance decays, the net baryon number, strangeness
and electric charge are conserved. Their event-by-event fluctuations
are expected to correspond to a grand canonical ensemble. 
In general, when dealing with fluctuations in QCD,
and in particular in relation to heavy ion collisions, it is important to
relate fluctuations of such conserved charges and the 
event-by-event fluctuations of observed (hadronic) species. The former have
been extensively studied with lattice simulations
\cite{Borsanyi:2011sw,Karsch:2012wm,Bazavov:2012jq,Bazavov:2012vg,Borsanyi:2013hza,Bellwied:2015lba,Ding:2015fca,Gunther:2016vcp,DElia:2016jqh,Borsanyi:2018grb},
and are essential to the study of the QCD phase diagram for multiple reasons.
First, they are directly related to the Taylor coefficients in the expansion of
the pressure to finite chemical potential and have been utilized to reconstruct
the Equation of State (EoS) of QCD at finite density, 
both in the case of sole baryon number conservation
\cite{Borsanyi:2012cr,Gunther:2016vcp,Bazavov:2017dus}, and with the inclusion
of all conserved charges \cite{Noronha-Hostler:2019ayj}. Second, higher order
fluctuations are expected to diverge as powers of the correlation length in the
vicinity of the critical point, and have thus been proposed as natural
signatures for its experimental search
\cite{Stephanov:1999zu,Stephanov:2011pb}. On the other hand, fluctuations of
observable particles can be measured in experiments, and are very closely
related to conserved charge fluctuations. With some caveats
\cite{Begun:2006jf,Kitazawa:2011wh,Kitazawa:2012at,Alba:2014eba}, comparisons
between the two can be made, provided that certain effects are taken into
account.

Previous studies found that, for certain particle species, fluctuations are
more sensitive to the freeze-out parameters than yields \cite{Alba:2015iva}.
In recent years, the STAR Collaboration has published results for the
fluctuations of net-proton \cite{Adamczyk:2013dal}, net-charge
\cite{Adamczyk:2014fia}, net-kaon \cite{Adamczyk:2017wsl}, and more recently
net-$\Lambda$ \cite{Nonaka:2019fhk} and for correlators between different
hadronic species \cite{Nonaka:2019fhk,Adam:2019xmk}. From the analysis of
net-proton and net-charge fluctuations in the HRG model, it was found that the
obtained freeze-out temperatures are lower than the corresponding ones from
fits of the yields \cite{Alba:2014eba}. More recent analyses
\cite{Bellwied:2018tkc,Bluhm:2018aei} of the moments of net-kaon distributions
showed that it is not possible to reproduce the experimental results for
net-kaon fluctuations with the same freeze-out parameters obtained from the
analysis of net-proton and net-charge. In particular, the obtained freeze-out
temperature is consistently higher, with a separation that increases with the
collision energy. In \cite{Bellwied:2018tkc}, predictions for the moments of
net-$\Lambda$ distributions were provided, calculated at the freeze-out of
net-kaons and net-proton/net-charge. 

Correlations between different conserved charges in QCD provide yet another
possibility for the comparison of theory and experiment. They will likely
receive further contribution from measurements in the future, with new species
being analyzed and increased statistics allowing for better determination of
moments of event-by-event distributions \cite{Adam:2019xmk}.

In this manuscript, we present continuum-extrapolated lattice QCD results for
all second order non-diagonal correlators of conserved charges. We then
identify the contribution of the single particle species to these correlators,
distinguishing between measured and non-measured species.  Finally, we identify
a set of observables, which can serve as proxies to measure the conserved
charge correlators. The manuscript is organized as follows. In Section
\ref{sec:lattice} we present the continuum extrapolated lattice results for
second-order non-diagonal correlators of conserved charges and discuss the
extrapolation to finite $\mu_B$ in Section \ref{sec:finitemu}.  In Section
\ref{sec:HRG2} we show the comparison with HRG model calculations, and describe
the breakdown of the different contributions to the observables shown in the
previous Section. In Section \ref{sec:proxies} we propose new observables which
can serve as proxies to directly study the correlation of conserved charges. In
Section \ref{sec:finite_mu_cuts} we analyze the behavior of the constructed
proxies at finite chemical potential, and study the
effect of acceptance cuts in the HRG calculations. We argue
that the small dependence on experimental effects allows for a direct comparison
with lattice QCD results. We also perform a comparison to experimental results
for selected observables in Section \ref{sec:exp_comp}. Finally, in Section \ref{sec:concl} we present our conclusions.

\section{Lattice QCD and the grand canonical ensemble\label{sec:lattice}}

The lattice formulation of quantum chromodynamics opens a non-perturbative
approach to the underlying quantum field theory in equilibrium. Its partition
function belongs to a grand canonical ensemble, parametrized by
the baryo-chemical potential $\mu_B$, the strangeness chemical potential
$\mu_S$ and the temperature $T$. Additional parameters include the volume
$L^3$, which is assumed to be large enough to have negligible volume effects,
and the quark masses. The latter control the pion and kaon masses, and are set
to reproduce their physical values. At the level of accuracy of this study we
can assume the degeneracy of the light quarks $m_u=m_d$ and neglect the effects
coming from quantum electrodynamics.  

There is a conserved charge corresponding to each flavor of QCD.
The grand canonical partition function can be then written in terms of
quark number chemical potentials ($\mu_u,\mu_d,\mu_s$).
The derivatives of the grand potential with respect to these chemical
potentials are the susceptibilities of quark flavors, defined as:
\begin{equation}
\chi^{u,d,s}_{i,j,k}= \frac{\partial^{i+j+k} (p/T^4)}{
(\partial \hat\mu_u)^i
(\partial \hat\mu_d)^j
(\partial \hat\mu_s)^k
} \, \, ,
\label{eq:chiquark}
\end{equation}
with $\hat\mu_q=\mu_q/T$. These derivatives are normalized to be dimensionless
and finite in the complete temperature range. 
For the purpose of phenomenology we introduce for the $B$ (baryon number),
$Q$ (electric charge) and $S$ (strangeness) a chemical potential
$\mu_B,\mu_Q$ and $\mu_S$, respectively.
The basis of $\mu_u,\mu_d,\mu_s$ can be transformed into a basis 
of $\mu_B,\mu_Q,\mu_S$ using the $B$, $Q$ and $S$ charges of the individual quarks:
\begin{eqnarray}
\mu_u&=&\frac13\mu_B+\frac23\mu_Q\,,\\
\mu_d&=&\frac13\mu_B-\frac13\mu_Q\,,\\
\mu_s&=&\frac13\mu_B-\frac13\mu_Q-\mu_S\,.
\end{eqnarray}

Susceptibilities are then defined as 
\begin{equation}
\chi^{BQS}_{ijk}(T, \hat{\mu}_B, \hat{\mu}_Q, \hat{\mu}_S) = \frac{\partial^{i+j+k} \left( p\left( T, \hat{\mu}_B, \hat{\mu}_Q, \hat{\mu}_S \right)/T^4 \right)}{\partial \hat{\mu}_B^i \partial \hat{\mu}_Q^j \partial \hat{\mu}_S^k} \, \, .
\label{eq:chiBQS}
\end{equation}

It is straightforward to express the derivatives of $p/T^4$ with 
respect to $\mu_B$, $\mu_Q$ and $\mu_S$ in terms of the coefficients in Eq.
(\ref{eq:chiquark}) \cite{Bernard:2004je,Bazavov:2012jq,Bellwied:2015lba}.
For the cross correlators we have

\begin{eqnarray}
\chi_{11}^{BQ}&=&\frac19 \left[
\chi^u_2-\chi^s_2-\chi_{11}^{us}+\chi_{11}^{ud}
\right]\,,\\
\chi_{11}^{BS}&=&-\frac13 \left[
\chi^s_2+2\chi_{11}^{us}
\right]\,,\\
\chi_{11}^{QS}&=&\frac13 \left[
\chi^s_2-\chi_{11}^{us}
\right]\,.\label{eq:chiQS}
\end{eqnarray}

Such derivatives play an important role in experiment. In an ideal
setup the mean of a conserved charge $i$ can be expressed as the first
derivative with respect to the chemical potential,
\begin{equation}
\langle N_i\rangle =  T \frac{\partial\log Z(T,V,\{\mu_q\})}{\partial\mu_i} \, \, ,
\end{equation}
while fluctuations and cross correlators (say between charges $i$ and $j$) are
second derivatives:
\begin{equation}
\frac{\partial\langle N_i\rangle}{\partial\mu_j}=
T \frac{\partial^2 \log Z(T,V,\{\mu_q\})}{\partial\mu_j\partial\mu_i}
=
\frac{1}{T}(\langle N_i N_j \rangle -\langle N_i  \rangle \langle N_j \rangle) \, \, .
\label{eq:fluctresponse}
\end{equation}
In these formulae $N_i$ indicates the net number of charge carriers, that is,
antiparticles come with an extra negative sign, e.g. number of
baryons - number of antibaryons for $B$.

The procedure to define the chemical potential on the lattice
\cite{Hasenfratz:1983ba} and to extract the derivatives in Eq. (\ref{eq:chiquark})
from simulations that run at $\mu_u=\mu_d=\mu_s=0$ has been worked out
long ago \cite{Allton:2002zi} and has been the basis of many studies
ever since \cite{Bernard:2004je,Gavai:2005yk,Allton:2005gk,Datta:2012pj}.
Since the derivatives with respect to the chemical potential require no
renormalization, a continuum limit could be computed as soon as results
on sufficiently fine lattices emerged \cite{Borsanyi:2010bp,Bazavov:2012jq}.
Later the temperature range and the accuracy of these extrapolations 
were extended in \cite{Bellwied:2015lba,Ding:2015fca}.

In this work, we extend our previous results \cite{Bellwied:2015lba} to
non-diagonal correlators and calculate a specific ratio that will be later
compared to experiment.

These expectation values are naturally volume dependent.
Their leading volume dependence can, however, be canceled by forming
ratios. In \cite{Bazavov:2012vg,Borsanyi:2013hza,Borsanyi:2014ewa}
such ratios were formed between various moments of electric charge
fluctuations, and also for baryon fluctuations. For the same ratios
the STAR experiment has provided proxies as part of the Beam Energy Scan I
program \cite{Adamczyk:2013dal,Adamczyk:2014fia}.

The gauge action is defined by the tree-level
Symanzik improvement, and the fermion action is a one-link staggered with
four levels of stout smearing. The parameters of the discretization as well as
the bare couplings and quark masses are given in \cite{Bellwied:2015lba}.

The charm quark is also included in our simulations, in order to account
for its partial pressure at temperatures above $200 \MeV$, where it is no longer
negligible \cite{Borsanyi:2016ksw}.  In the range of the expected chemical
freeze-out temperature between $135$ and $165 \MeV$ the effect of the charm
quark is not noticeable on the lighter flavors \cite{Bellwied:2015lba}.

In this work we use the lattice sizes of $32^3\times 8$, $40^3\times10$,
$48^3\times12$, $64^3\times16$ as well as $80^3\times20$. Thus, the physical
volume $L^3$ is given in terms of the temperature as $LT=4$ throughout this
paper.  The coarsest lattice was never in the scaling region. The finest
lattice lacks the precision of the others, and we only use it when the coarser
lattices, e.g. $40^3\times10$, are not well in the scaling region and if, for a
particular observable, the $80^3\times20$ has small enough error bars. If this
was used, the data set is also shown in the plots.

\begin{figure}
\includegraphics[width=0.99\linewidth]{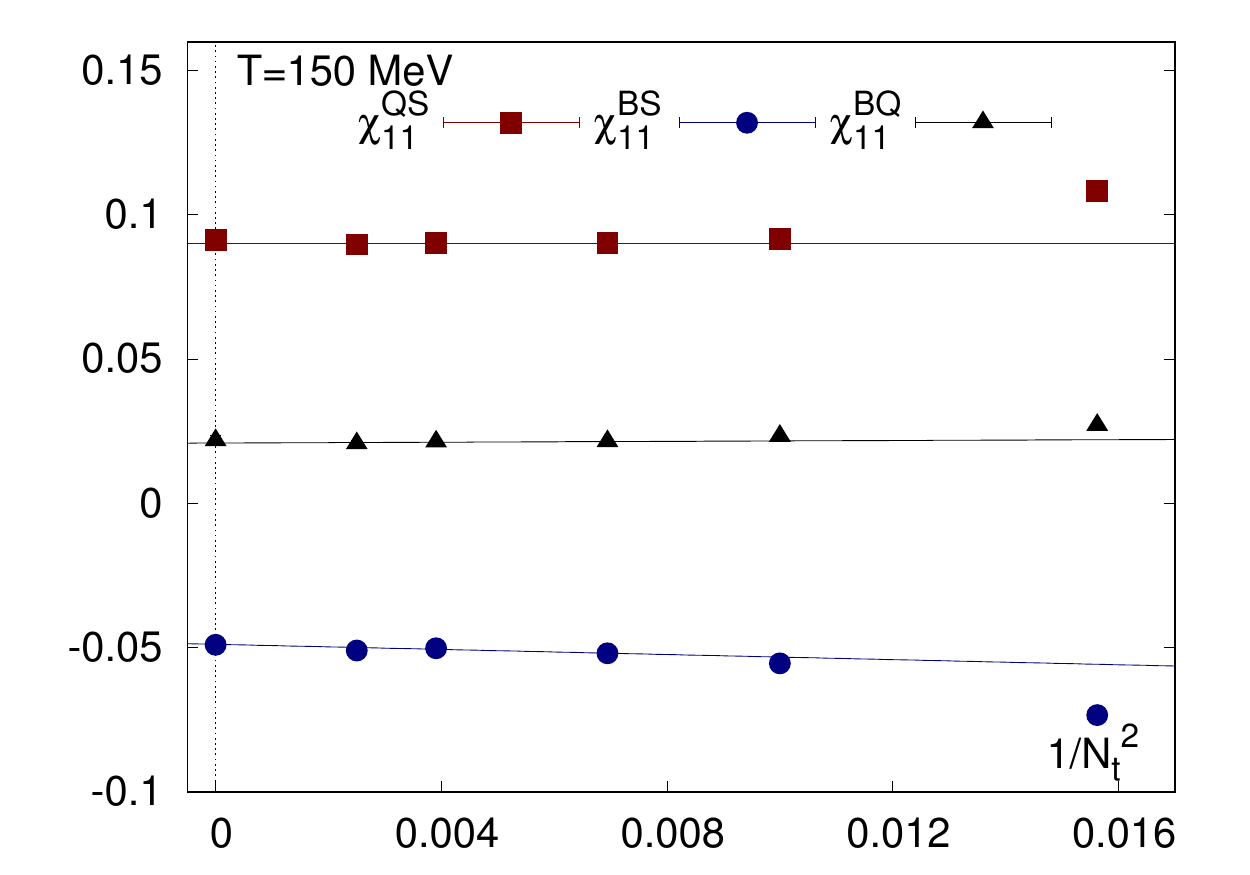}
\caption{\label{fig:cont_ext_example}
Examples for the continuum extrapolation. We show the three
cross correlators on the lattices (from right to left): 
$32^3\times8$, $40^3\times10$, $48^3\times12$, $64^3\times16$ and
$80^3\times20$.
The data points correspond to the $w_0$-based scale setting \cite{Borsanyi:2012zs},
one of the two interpolation methods to align all simulation results
to the same temperature: $T = 150 \MeV$ in this example. The error bars in the
continuum limit are obtained from the combination of the scale setting,
the interpolation, the selection of the continuum extrapolation fit range,
and whether a linear or 1/linear function is fitted.
}.
\end{figure}

\begin{figure}
\includegraphics[width=0.99\linewidth]{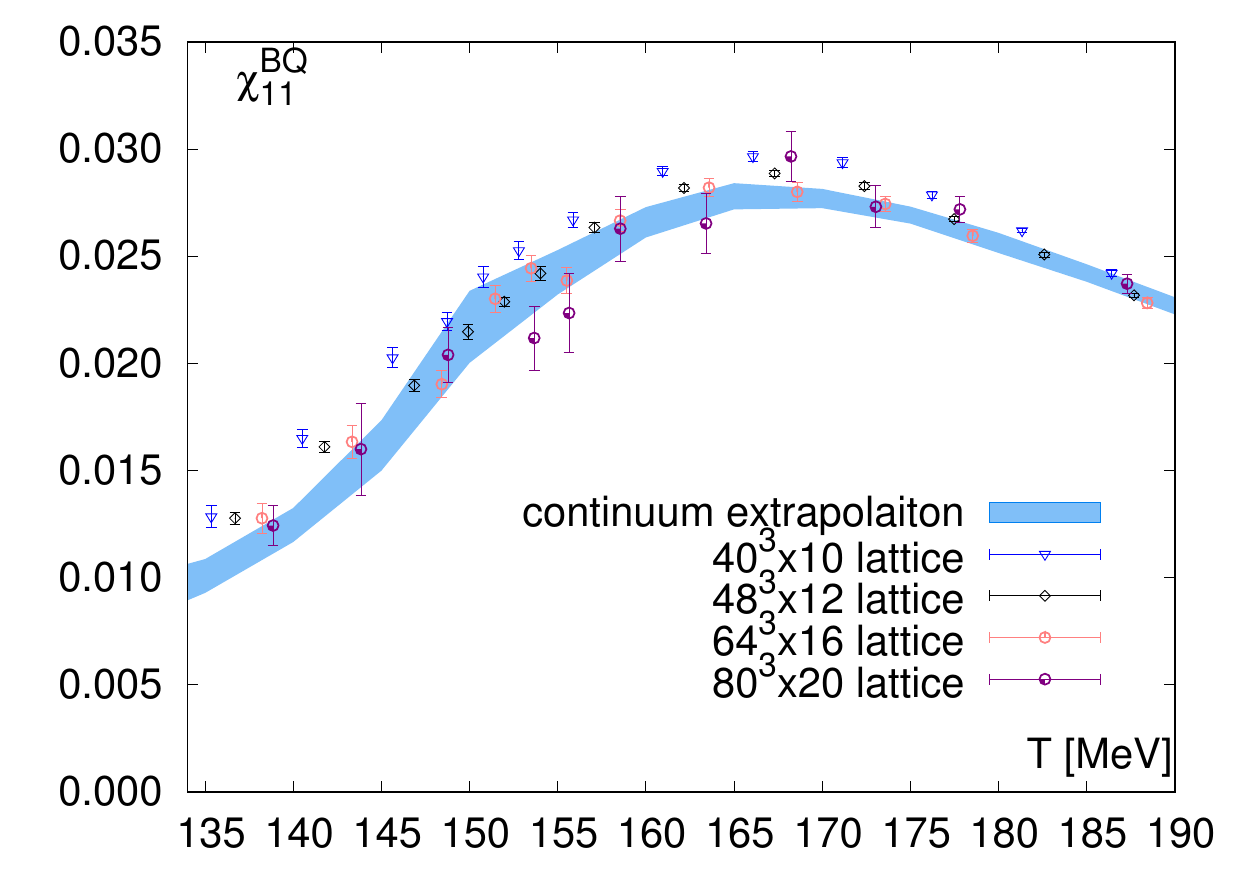}
\caption{\label{fig:BQ}
The baryon-electric charge cross correlator from the lattice at finite lattice spacing and
its continuum limit. 
}. 
\end{figure}

We show here the continuum extrapolated cross correlators at zero
chemical potential. In Fig. \ref{fig:cont_ext_example} we show an example
of continuum extrapolation for the three cross correlators, with $T= 150 \MeV$ 
and the $w_0$-based scale setting \cite{Borsanyi:2012zs}.  Fig.~\ref{fig:BQ} 
shows $\chi^{BQ}_{11}(T)$ for the four different lattices, as well as the 
continuum extrapolation. Although our simulation contains a dynamical charm 
quark, we did not account for its baryon charge. Thus, the Stefan-Boltzmann limit
of this quantity is zero. This limit is reached when the mass difference
between the strange and light quarks becomes negligible in comparison to
the temperature. The peak is seen at a higher temperature than $T_c\approx 155 \MeV$,
while in the transition region there is an inflection point. Below $T_c$ this
correlator is dominated by protons and charged hyperons. In Section \ref{sec:HRG2} we will account in detail for various hadronic
contributions in the confined phase.

\begin{figure}
\includegraphics[width=0.99\linewidth]{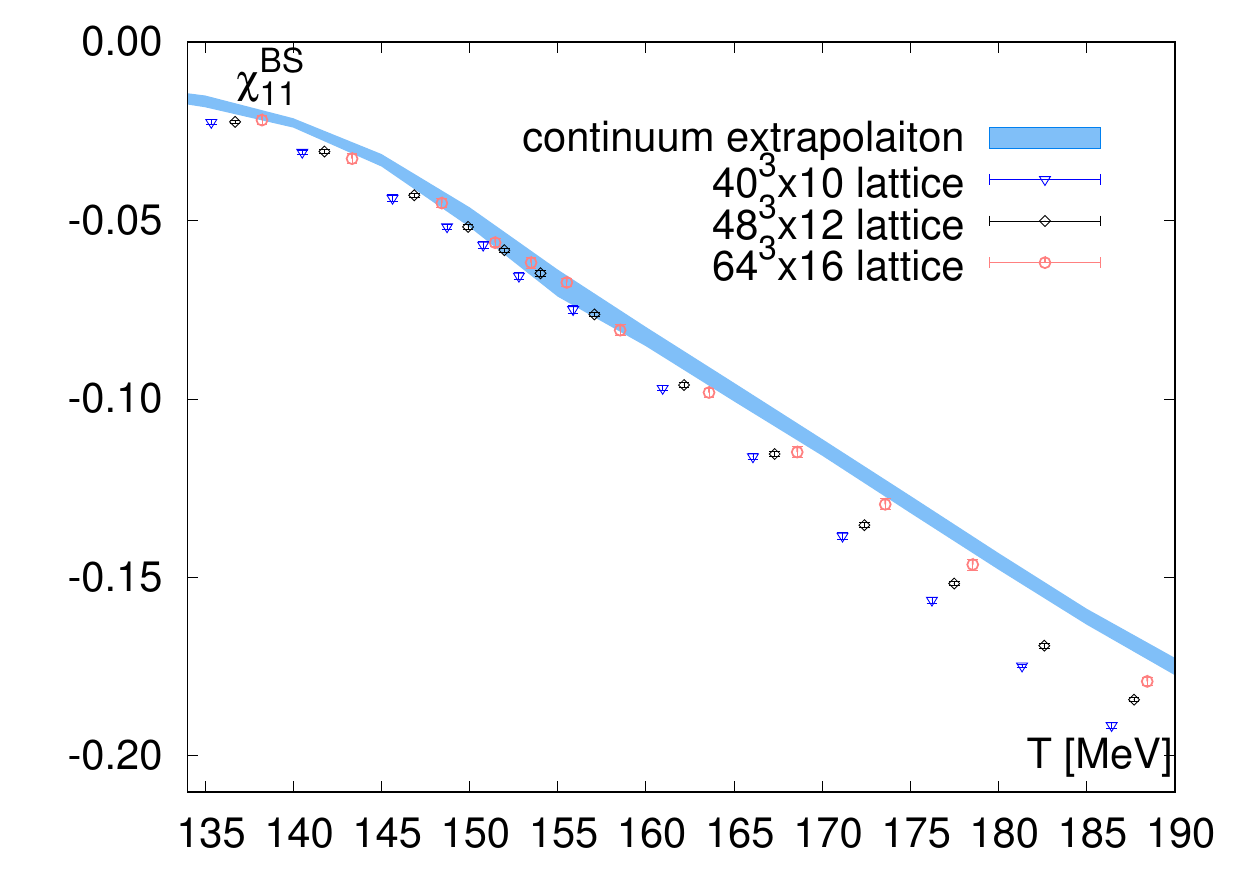}
\caption{\label{fig:BS}
The baryon-strangeness cross correlator from the lattice at finite lattice spacing and
its continuum limit.
}.
\end{figure}

The $\chi^{BS}_{11}(T)$ correlator is shown in Fig.~\ref{fig:BS}.
Unlike the $BQ$ correlator, we have now a monotonic function
with a high temperature limit of $-1/3$, which is the baryon
number of the strange quark. The transition has a remarkably
small effect on this quantity. At low temperatures this correlator
is basically the hyperon free energy.

\begin{figure}
\includegraphics[width=0.99\linewidth]{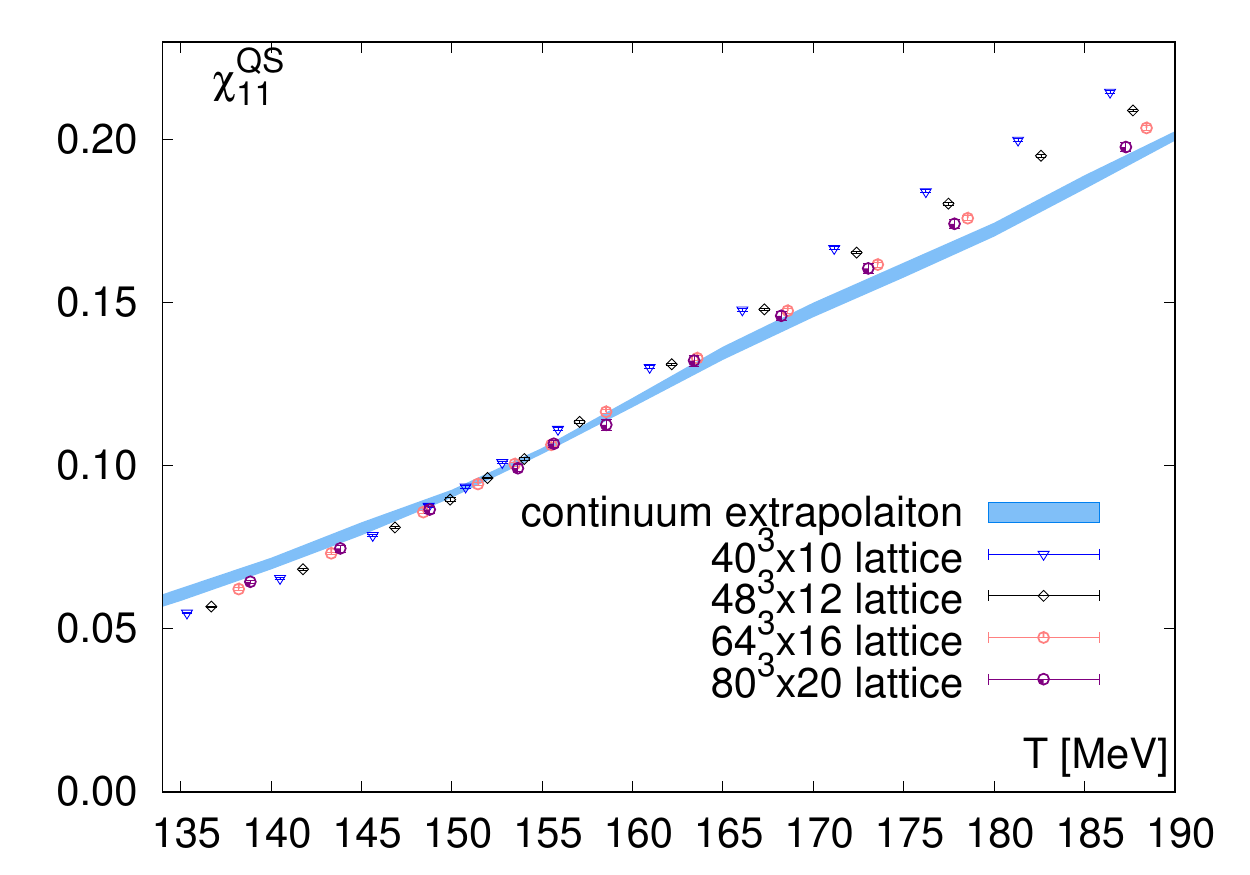}
\caption{\label{fig:QS}
The electric charge-strangeness cross correlator from the lattice at finite lattice spacing and its continuum limit.
}.
\end{figure}

The $\chi^{QS}_{11}(T)$ correlator in Fig.~\ref{fig:QS}
is also monotonic, converging to $1/3$ at high $T$, which is the
electric charge of the strange quark. At low temperatures this quantity
is dominated by the charged kaons, which were in the focus
of recent experimental investigations \cite{Adamczyk:2017wsl}.

\section{Results at finite density\label{sec:finitemu}}

Since lattice QCD can be defined at finite values of the $B$, $Q$ and $S$
chemical potentials, and is capable of calculating derivatives of the
free energy as a function of these chemical potentials one could expect
that the extension of the simulations to finite density is a mere technical
detail. Unfortunately, at any finite real value of the quark chemical potential
$\mu_q$ the fermionic contribution to the action becomes complex and
most simulation algorithms break down. 

There are several options to extract physics at finite densities,
nevertheless. It seems natural to use algorithms that were designed to
work on complex actions -- both the complex Langevin equation
\cite{Sexty:2019vqx} and the Lefschetz thimble approach \cite{Scorzato:2015qts}
have shown promising results recently -- yet their direct application to
phenomenology requires further research.

Instead, we use here the parameter domain that is available for
main stream lattice simulations. In fact, besides zero chemical potential,
simulations at imaginary $\mu_B$ are also possible, and have been exploited
in the past to extrapolate the transition temperature
\cite{Bonati:2015bha,Bellwied:2015rza,Cea:2015cya,Bonati:2018nut},
fluctuations of conserved charges 
\cite{DElia:2016jqh,Borsanyi:2018grb} and the equation of state \cite{Gunther:2016vcp}. 
In all these works it was assumed that the thermodynamical observables
are all analytical functions of $\hmu_B^2$.

A conceptually very similar method, the Taylor method, provides the
extrapolation in terms of calculating higher derivatives with respect to
$\hmu_B$. The series is truncated at a certain order, which is typically
limited by the statistics of the lattice simulation.

Since we will relate the baryon-strangeness correlator to experimental
observables later on we use $\chi^{BS}_{11}$ as an example for the Taylor expansion:
\begin{eqnarray}
\chi^{BS}_{11} (T, \hmu) &=&
\chi^{BS}_{11} (T, 0)  
+ \hmu_B \chi^{BS}_{21} (T, 0) 
+ \hmu_S \chi^{BS}_{12} (T, 0) \nonumber\\
&&
+ \frac{\hmu_B^2}{2} \chi^{BS}_{31} (T, 0) 
+ \hmu_B\hmu_S\chi^{BS}_{22} (T, 0) \nonumber\\
&&
+ \frac{\hmu_S^2}{2} \chi^{BS}_{13} (T, 0) 
+ \mathcal{O}(\hmu^4) \, \, ,
\end{eqnarray}
where the terms proportional to $\hmu_B$ and $\hmu_S$ vanish since
they contain odd derivatives, which are forbidden by the  $C$-symmetry of QCD.
(The charge chemical potential is omitted for simplicity.)

In most phenomenological lattice studies the chemical potentials are selected
such that the strangeness vanishes for each set of $(\mu_B,\mu_Q,\mu_S)$.
More precisely: for each $T$ and $\hmu_B$ we select $\hmu_Q$ and $\hmu_S$
values such that 
\begin{eqnarray}
\chi^S_1(T,\{\hmu_i\}) &=& 0 \, \, ,\nonumber\\
\chi^Q_1(T,\{\hmu_i\}) &=& 
0.4 \chi^B_1(T,\{\hmu_i\}) \, \, .
\label{eq:chineutrality}
\end{eqnarray}
The factor 0.4 is the typical $Z/A$ ratio for the projectiles in the heavy ion
collision setup, and the value we use in the HRG model calculations below. In our lattice 
study, however, we will use 0.5. This
introduces a small effect compared to the statistical and systematic errors
of the extrapolation, and results in substantial simplification of the
formalism: $\mu_Q$ can be chosen to be 0. The would-be $\mu_Q$ value is about one
tenth of $\mu_S$ in the transition region \cite{Bazavov:2012vg,Borsanyi:2013hza}. We
checked the impact of our simplification on the results we present here, with the Taylor 
expansion method. Utilizing our simulations data from ensembles at $\mu_B=0$, we 
calculate the correction to the ratio $\chi^{BS}_{11}/\chi^{S}_2$. By construction, this 
correction vanishes 
at $\mu_B=0$, and we find that it grows to at most $1\%$ at $\mu_B/T=1$, and at most 
$1.5\%$ at $\mu_B/T=2$. These systematic errors are considerably smaller than the 
uncertainties we have on our results, as can be seen in Fig.~\ref{fig:taylorvssector}.

\subsection{Taylor method}

The Taylor coefficients for correlators can be easily obtained by
considering the higher derivatives with respect to $\mu_B$.
For later reference we select the quantity $\chi^{BS}_{11}/\chi^{S}_2$
for closer inspection:
\begin{equation}
\left.\frac{\chi^{BS}_{11}}{\chi^{S}_2}\right|_{\mu_B/T} 
=
\frac{\chi^{BS}_{11}}{\chi^{S}_2}  + 
\frac{\hmu_B^2}{2}
\frac{\chi^{BS,(NLO)}_{11}\chi^{S}_2-\chi^{S,(NLO)}_2\chi^{BS}_{11}
}{(\chi^{S}_2)^2}\, \, ,
\end{equation}
up to $\mathcal{O}(\hmu_B^4)$ corrections, with
\begin{eqnarray}
\chi^{BS,(NLO)}_{11}&=&\chi^{BS}_{13} s_1^2 + 2\chi^{BS}_{22} s_1 + \chi^{BS}_{31} \, \, ,\\
\chi^{S,(NLO)}_{2}&=&\chi^{BS}_{22} + 2\chi^{BS}_{13}s_1 + \chi^{S}_4 s_1^2 \, \, , \\
s_1&=& - \chi^{BS}_{11}/\chi^{S}_2\, \, .
\end{eqnarray}
The derivatives on the right hand side are all taken at $\mu_B=\mu_S=0$.

Whether we extract the required derivatives from a single simulation (one per
temperature) at $\mu_B=\mu_S=0$, or we determine the fourth order derivatives
numerically by the (imaginary) $\mu_B$-dependence of second order derivatives
is a result of a cost benefit analysis. The equivalence of these two choices
has been shown on simulation data (of the chemical potential dependence
of the transition temperature) in \cite{Bonati:2018nut}.

\begin{figure}
\includegraphics[width=0.99\linewidth]{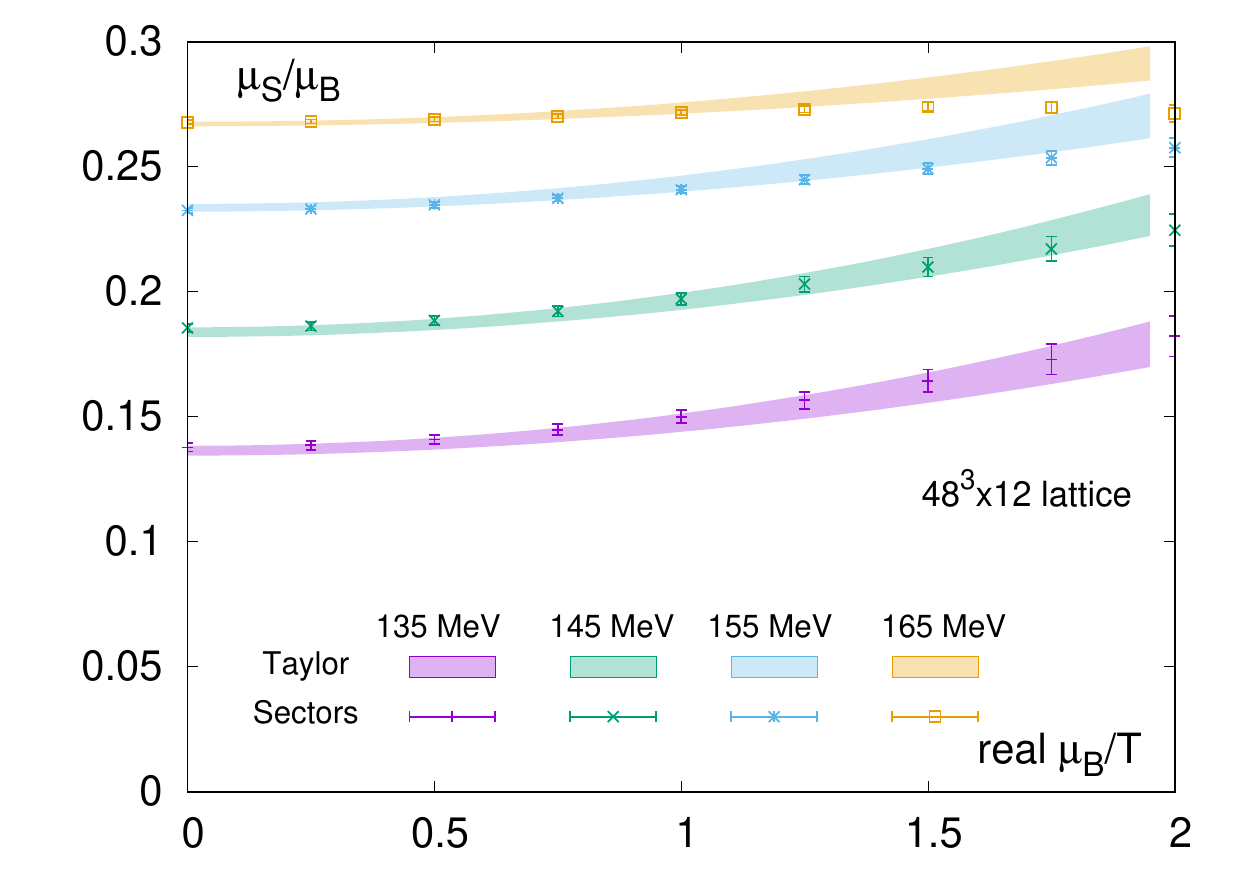}
\includegraphics[width=0.99\linewidth]{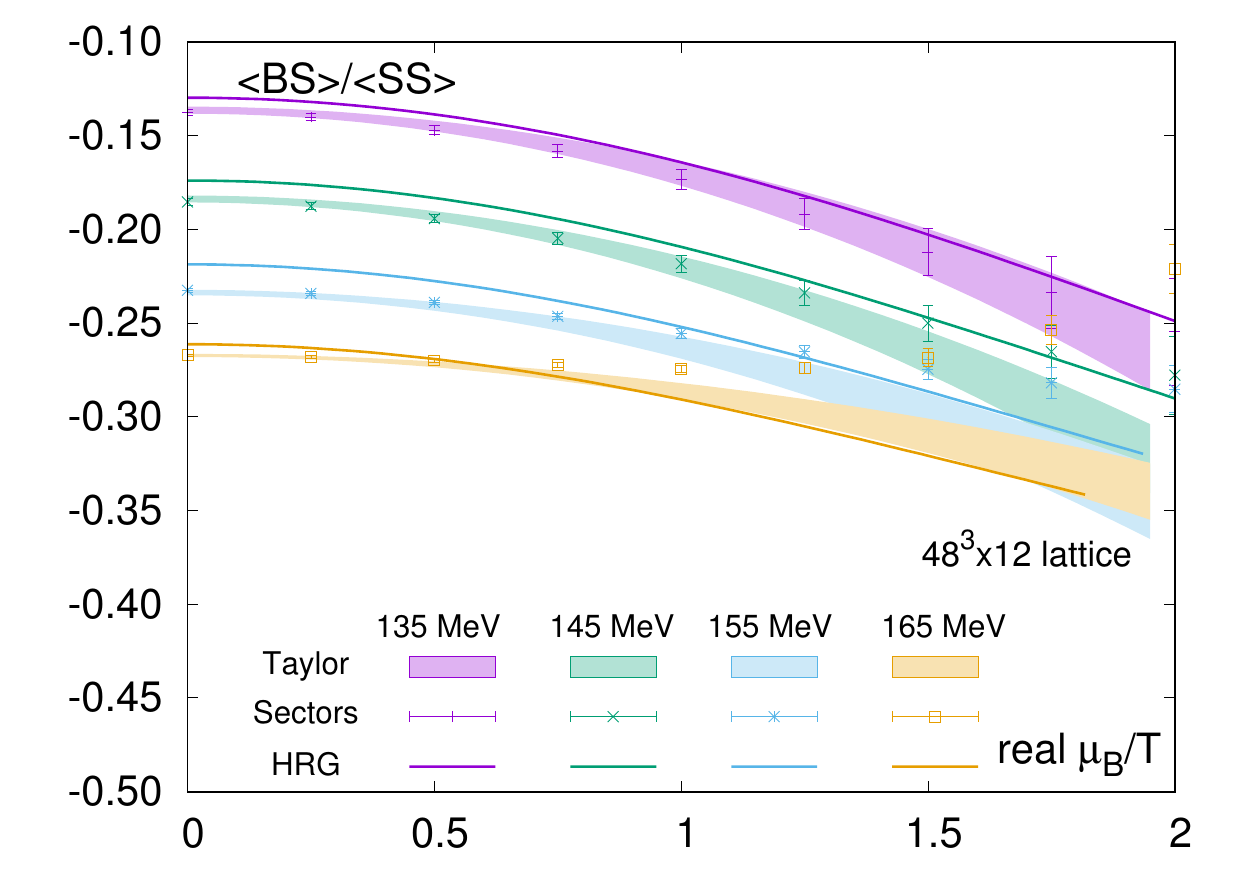}
\caption{\label{fig:taylorvssector}
Comparison of two approaches to the finite density extrapolation of two
observables. 
The Taylor result is truncated such that only the leading $\sim \hmu_B^2$
contribution is considered. In the sector method, contributions up to
$|B|=2$ are included. (The data were generated from a $48^3\times12$
lattice, the plots show the intermediate result before continuum
extrapolation.) We also show the Hadron Resonance Gas model prediction.
}
\end{figure}

In \cite{Bellwied:2015rza} we calculated direct derivatives; however,
we obtained smaller errors by using imaginary $\mu_B$ simulations in
\cite{Borsanyi:2018grb}. Thus, we take the Taylor coefficients from
the latter analysis, now extended to the new observable. The results for
several fixed temperatures are shown in Fig.~\ref{fig:taylorvssector}. 
As a first observable we show the $\mu_S/\mu_B$ ratio that realizes
strangeness neutrality. Then we show the $\chi^{BS}_{11}/\chi^{S}_2$ ratio
as a function of positive $\hmu_B^2$.

In the plots we show results from a specific lattice $48^3\times12$, which has
the highest statistics, so that it provides the best ground to compare different
extrapolation strategies.

\subsection{Sector method\label{sec:sectors}}

In Fig.~\ref{fig:taylorvssector} we compare two extrapolation strategies,
here we describe the second approach, the sector method.

We are building on our earlier work in \cite{Alba:2017mqu}, where
we have written the pressure of QCD as a sum of the sectors
\begin{eqnarray}
\label{eq:pressure}
P(\hmu_B,\hmu_S) &=& P^{BS}_{00}+P^{BS}_{10}\cosh(\hmu_B)+P^{BS}_{01} \cosh(\hmu_S)
\nonumber \\
&+& P^{BS}_{11} \cosh(\hmu_B-\hmu_S)
\nonumber \\
&+& P^{BS}_{12} \cosh(\hmu_B-2\hmu_S)
\nonumber \\
&+& P^{BS}_{13} \cosh(\hmu_B-3\hmu_S)
\;.
\end{eqnarray}
These sectors were also studied in \cite{Bazavov:2013dta,Noronha-Hostler:2016rpd}. Obviously, QCD receives contributions from sectors with higher quantum numbers as well. 
The sectors in Eq. (\ref{eq:pressure}) are the only ones receiving contributions from the
ideal Hadron Resonance Gas model in the Boltzmann approximation.  (The
dependence on the electric charge chemical potential is not considered now, 
since we selected $\mu_Q=0$.)

The partitioning of the QCD pressure in sectors is very natural
in the space of imaginary chemical potentials $\hmu_B=i\hmu_B^I$ and
$\hmu_S=i\hmu_S^I$:
\begin{equation}
P(\hmu_B^I,\hmu_S^I) =\sum_{j,k} P^{BS}_{jk} \cos(j\hmu_B^I-k\hmu_S^I) \, \, .
\end{equation}
It is expected that higher sectors will be increasingly relevant
as $T_c$ is approached from below. A study using Wuppertal-Budapest
simulation data has shown that below $T \approx 165 \MeV$
the sectors $|B|=0,1,2$ give a reasonable description, e.g. by calculating
$\chi^{B}_4$ from the sectors coefficients and comparing to direct results.

Thus, for this work we considered the next-to-leading order of the
sector expansion, including the
\begin{eqnarray}\label{eq:usedsectors}
 LO:&& \quad
P^{BS}_{01},~ P^{BS}_{11},~ P^{BS}_{12},~ P^{BS}_{13} \, \, , \nonumber\\
 NLO:&& \quad
P^{BS}_{02},~ P^{BS}_{1,-1},~ P^{BS}_{21},~ P^{BS}_{22} \, \, .
\end{eqnarray}

It is somewhat ambiguous how the $NLO$ is to be defined. One option would be to
include the next-higher $|B|$ quantum number, making our approach 2nd order in
this expansion. We do include $P^{BS}_{21}$ and $ P^{BS}_{22}$, however adding
further higher strangeness sectors -- e.g. $P^{BS}_{23}$ -- did not 
improve the agreement with our data, so it was not included. Moreover, we included the
multi-strange sector $P^{BS}_{02}$ and the exotic sector $ P^{BS}_{1,-1}$, which is the 
coefficient of the term proportional to $\cosh (\hat{\mu}_B + \hat{\mu}_S)$. On the other 
hand,removing the sectors included in the $NLO$ never resulted at higher
temperatures in a smaller $\chi^2$ for the fit, e.g. at $T=160 \MeV$
removing the terms $P^{BS}_{0,2}$, $P^{BS}_{1,-1}$  $P^{BS}_{2,1}$ or
$P^{BS}_{2,0}$  from the fit (removing only one term at a time) resulted in a
$\chi^2/\mathrm{N}_\mathrm{dof}$ of $72.6/53$, $181.1/53$, $72.4/53$, or $137.2/53$, 
respectively, while including all gave $72.3/52$.

\begin{figure}
\includegraphics[width=0.99\linewidth]{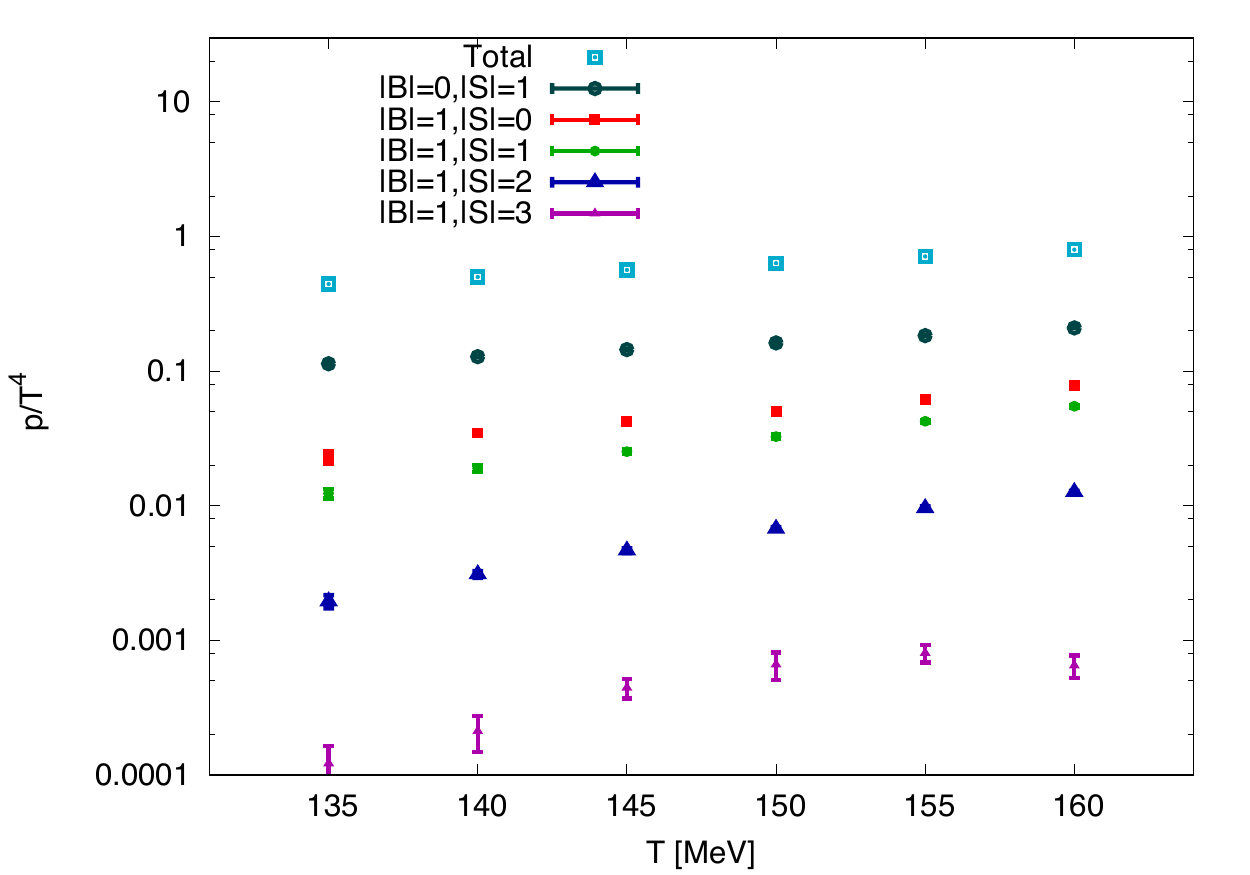}
\includegraphics[width=0.99\linewidth]{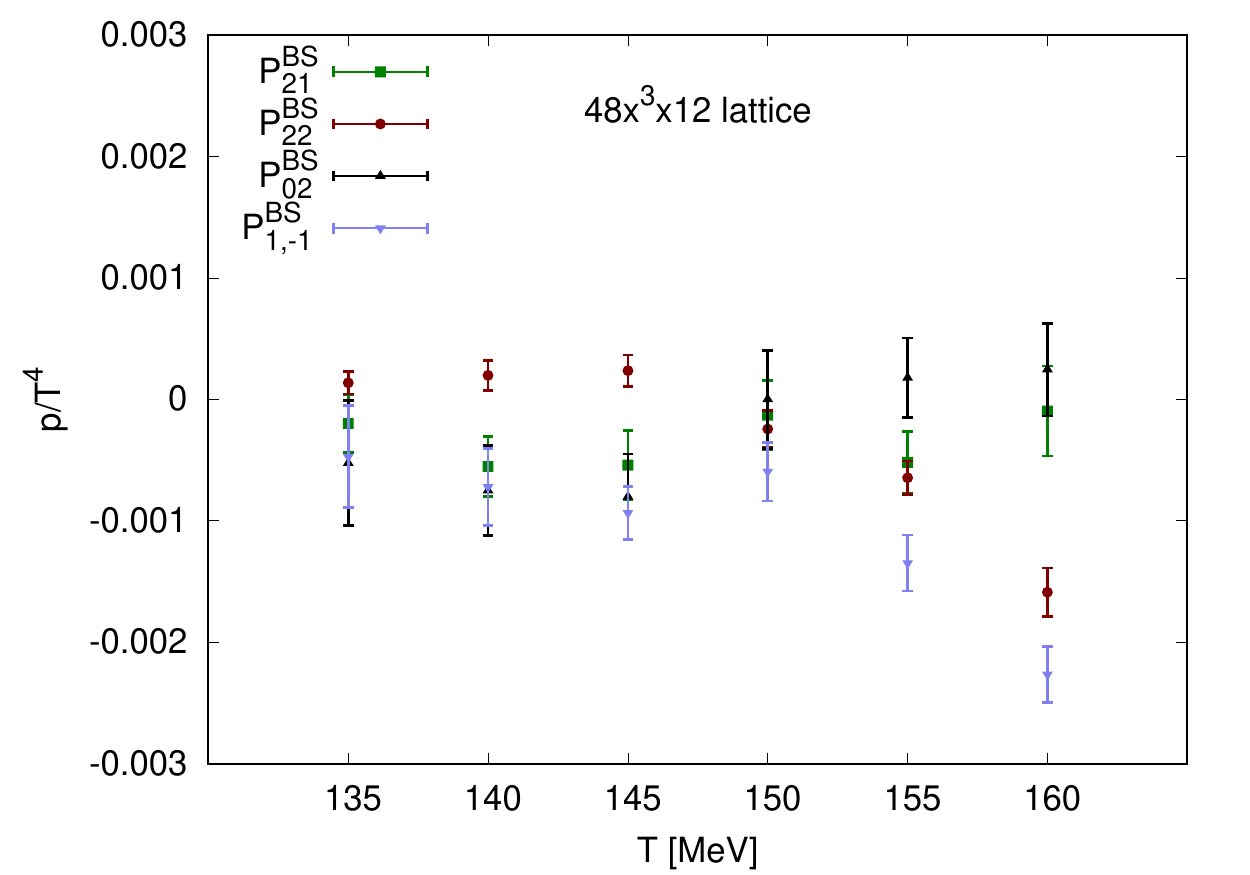}
\caption{\label{fig:sectors}
The magnitude of the various sector coefficients in the temperature
region relevant for freeze-out studies. In the first panel we
show the standard sectors on a logarithmic scale as published in our earlier work
\cite{Alba:2017mqu}.  In the second panel we show the non-standard sectors in a
linear scale that we use for the $\mu_B$-extrapolation in this work. Note that results from the same lattice for the $|B|$-only sectors are shown in \cite{Vovchenko:2017xad}, and can be taken as a comparison.
}
\end{figure}

In Fig.~\ref{fig:sectors} the results for the sectors in the $LO$ and $NLO$ are shown at 
different temperatures. The results for the $LO$ sectors shown in the top panel (figure 
from Ref.~\cite{Alba:2017mqu}) are continuum extrapolated -- except for the 
$| B | = 0$, $| S | = 1$ sector. In the lower panel new results for the sectors in the $NLO$ 
are shown, generated from a $48^3\times12$ lattice. 

The alert reader may ask why we do not include the $P^{BS}_{10}$ sector,
accounting e.g. for protons. In fact, the sectors with $|S|=0$ do not
contribute to the observables $\chi^{BS}_{11}$, $\chi^{S}_2$ or $\chi^{S}_1$
neither at zero, nor at any real or imaginary chemical potential. 
In the analysis we included results for $\chi^{S}_1$, $\chi^{S}_2$, $\chi^{BS}_{11}$ from various 
data sets at various imaginary chemical potentials:
at $\mu_B=0,\mu_S=0$ the data of Section \ref{sec:lattice};
the $\mu_B^I>0,\mu_S=0$ data set of \cite{Borsanyi:2018grb};
the strangeness data set with $\mu_B^I>0$ of \cite{Gunther:2016vcp};
and finally the set (only including $\chi^S_1$ and $\chi^S_2$) with  $\mu_B=0,\mu_S^I>0$
from \cite{Alba:2017mqu}.
For the lower temperatures the model defined with the coefficients in Eq.
(\ref{eq:usedsectors}) resulted in good fits ($Q$ values ranging from 0 to 1) --
the worst fit was at $T=165 \MeV$ with $Q\approx 0.05$. This is the
temperature where the model is expected to break down.

Now we can compare the results to the Taylor expansion.
In Fig.~\ref{fig:taylorvssector} we show the sector results with error
bars, while the bands refer to the Taylor method. At low temperatures we see 
good agreement even for large values of the chemical potential; near the transition, 
however, the two approaches deviate already in the
experimentally relevant region. It is obvious that the sector
method breaks down above $T_c$. Its systematic improvement to higher
$|B|$ quantum numbers requires much higher statistics (the same is true
for the Taylor coefficients). Each further order enables the extrapolation
to somewhat larger chemical potential, and in the case of the $|B|$ sectors,
to a somewhat larger temperature. Let us note that the Taylor method
has limitations as well, slightly above $T_c$, because the subsequent orders are not
getting smaller \cite{Borsanyi:2018grb}. The reason for this behaviour is the
fact that, between $T = 160 \MeV$ and $T = 180 \MeV$, there is a cross-over 
transition in the imaginary domain of $\hmu_B$, then higher Taylor coefficients 
facilitate an extrapolation through that crossover.

\begin{figure}
\includegraphics[width=0.99\linewidth]{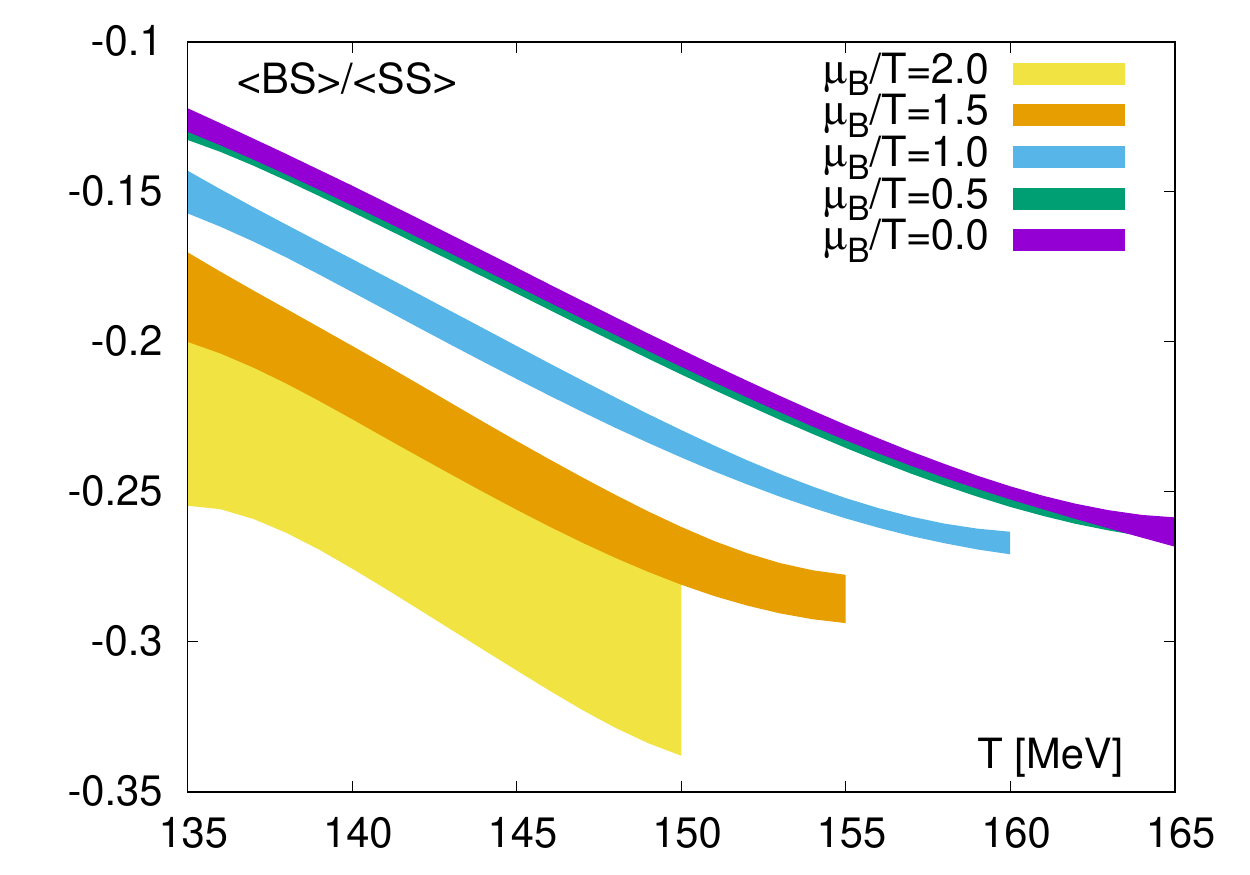}
\caption{\label{fig:rbs_cont}
Continuum extrapolation of the $\chi^{BS}_{11}/\chi^S_2$ ratio
as a function of the temperature for selected fixed real $\hmu_B$ values, 
obtained using the sector method.
}
\end{figure}

In conclusion, we consider only the chemical potential range where
our two methods agree in the extrapolation. At present, our lattice data
allow a continuum extrapolation from the sector method only, which we
do using $40^3\times10$, $48^3\times12$ and $64^3\times16$ lattices
in the temperature range $135 - 165 \MeV$ for a selection of fixed real $\hmu_B$ 
values. The method for the continuum extrapolation is the same we also used in 
Section \ref{sec:lattice}. We show the result in Fig.~\ref{fig:rbs_cont}.

The large error bars in comparison to the $\mu_B=0$ results
and the limited range in $\hmu_B$ indicate that the extraction of finite
density physics from $\mu_B=0$ or imaginary $\mu_B$ simulations
is a highly non-trivial task. Still, both the Taylor and the sector methods
can be systematically improved to cover more of the range of interest for the
Beam Energy Scan II program. Given the high chemical freeze-out temperatures 
for $|S| = 1$ particles -- see Section \ref{sec:exp_comp} -- that emerge from
STAR data (\cite{Adamczyk:2017wsl} and preliminary \cite{Nonaka:2019fhk}), 
the use of continuum extrapolated lattice simulations to
calculate the grand canonical features of QCD is highly motivated.

\section{Correlators in the HRG model} \label{sec:HRG2}

The HRG model is based on the idea that a gas of interacting hadrons in their ground state can be well described by a gas of non-interacting hadrons and resonances. The partition function of the model can thus be written as a sum of ideal gas contributions of all known hadronic resonances $R$:
\begin{equation}
\frac{p}{T^4} = \frac{1}{T^4} \sum_R p_R = \frac{1}{VT^3} \sum_R \ln \calZ_R (T, \vec{\mu}) \, \, ,
\end{equation}
with:
\begin{equation}
\ln \calZ_R = \eta_R \frac{V d_R}{2 \pi^2 T^3} \bigintssss_0^\infty \fourback dp \, p^2 \log \left[ 1 - \eta_R z_R \exp \left( - \epsilon_R/T \right) \right] \, \, ,
\end{equation}
where every quantity with a subscript $R$ depends on the specific particle in the sum. The relativistic energy is $\epsilon_R = \sqrt{p^2 + m_R^2}$, the fugacity is $z_R = \exp\left( \mu_R /T \right)$, the chemical potential associated to $R$ is $\mu_R = \mu_B B_R + \mu_Q Q_R + \mu_S S_R$, the conserved charges $B_R$, $Q_R$ and $S_R$ are the baryon number, electric charge and strangeness respectively.  Moreover, $d_R$ is the spin degeneracy, $m_R$ the mass, and the factor $\eta_R = \left(-1 \right)^{1+B_R}$ is $1$ for (anti)baryons and $-1$ for mesons.
 
The temperature and the three chemical potentials are not independent, as 
the conditions in Eqs.~(\ref{eq:chineutrality}) 
are imposed on the baryon, electric charge and strangeness densities.
We use these constraints to set both $\mu_Q(T,\mu_B)$ and $\mu_S(T,\mu_B)$
in our HRG model calculations.

In this work we utilize the hadron list PDG2016+ from \cite{Alba:2017mqu}, which was constructed with all the hadronic states (with the exclusion of charm and bottom quarks) listed by the Particle Data Group (PDG), including the less-established states labeled by $^*,^{**}$ \cite{Patrignani:2016xqp}. The decay properties of the states in the list, when not available (or complete) from the PDG, were completed with a procedure explained in \cite{Alba:2017hhe}, and then utilized in \cite{Alba:2017hhe,Bellwied:2018tkc}.
 
In the HRG model the $\chi^{BQS}_{ijk}$
susceptibilities of Eq.~(\ref{eq:chiBQS}) can be expressed as:
\begin{equation}
\chi^{BQS}_{ijk} \left(T, \hat{\mu}_B, \hat{\mu}_Q, \hat{\mu}_S \right) = \sum_R B_R^i \, Q_R^j \, S_R^k \,  I^R_{ijk} \left(T, \hat{\mu}_B, \hat{\mu}_Q, \hat{\mu}_S \right) \, \, ,
\end{equation}
where $B_R,Q_R,S_R$ are the baryon number, electric charge and strangeness of the species $R$ and the phase space integral at order $i+j+k$ reads (note that it is completely symmetric in all indices, hence $i+j+k = l$):
\begin{equation}
I^R_{l} \left(T, \hat{\mu}_B, \hat{\mu}_Q, \hat{\mu}_S \right) = \frac{\partial^{l} p_R/T^4}{\partial \hat{\mu}_R^{l}} \, \, .
\end{equation}

The HRG model has the advantage, when comparing to experiment, of allowing for the inclusion of acceptance cuts and resonance decay feed-down, which cannot be taken into account in lattice QCD calculations.  

The acceptance cuts on transverse momentum and rapidity (or pseudorapidity) can be easily taken into account in the phase space integrations via the change(s) of variables:
\begin{align}
\frac{1}{2 \pi^2}\bigintssss_0^\infty \!\!\!\!\!\! dp \, p^2 \rightarrow& \frac{1}{4 \pi^2} \bigintssss_{y^A}^{y^B} \!\!\!\!\!\! dy \bigintssss_{p_T^A}^{p_T^B} \!\!\!\!\!\! dp_T  \, p_T \cosh y  \sqrt{p_T^2 + m^2} \\ \nonumber
\rightarrow& \frac{1}{4 \pi^2} \bigintssss_{\eta^A}^{\eta^B} \!\!\!\!\!\! d\eta \bigintssss_{p_T^A}^{p_T^B} \!\!\!\!\!\! dp_T  \, p_T^2 \cosh \eta
\end{align}
in the case of rapidity and pseudorapidity respectively, where in all cases the trivial angular integrals were carried out \cite{Alba:2014eba}.

\subsection{Correlators of measured particle species \label{sec:breakdown_generic}}

The rich information contained in the system created in a heavy ion collision about the correlations between conserved charges is eventually carried over to the final stages through hadronic species correlations and self-correlations. It is convenient, in the framework of the HRG model, to consider the hadronic species which are stable under strong interactions, as these are the observable states accessible to experiment. However, due to experimental limitations, charged particles and lighter particles are easier to measure, and so we cannot access every relevant hadron related to conserved charges. Thus, historically protons have served as a proxy for baryon number, kaons as a proxy for strangeness, and net electric charge is measured through $p$, $\pi$, and $K$.

In our framework, we consider the following species, stable under strong interactions: $\pi^0$, $\pi^\pm$, $K^\pm$, $K^0$, $\overline{K}^0$, $p$,  $\overline{p}$, $n$, $\overline{n}$ , $\Lambda$, $\overline{\Lambda}$, $\Sigma^+$, $\overline{\Sigma}^-$, $\Sigma^-$, $\overline{\Sigma}^+$, $\Xi^0$, $\overline{\Xi}^0$, $\Xi^-$, $\overline{\Xi}^+$, $\Omega^-$,  $\overline{\Omega}^+$. Of these, the commonly measured ones are the following:
$$\pi^\pm , \, \,  K^\pm , \, \,  p \left( \overline{p}\right) , \, \,  \Lambda ( \overline{\Lambda}) , \, \,  \Xi^- ( \overline{\Xi}^+) , \, \,  \Omega^- ( \overline{\Omega}^+) .$$

A few remarks are in order here. First of all, we refer to the listed species as commonly measured because, although some others are potentially measurable (especially the charged $\Sigma$ baryons), results for their yields or fluctuations are not routinely performed both at RHIC and the LHC. In the following, we will keep our nomenclature of ``measured'' and ``non-measured" in accordance to the separation we adopt here. Obviously, neutral pions can be measured with the process $\pi^0 \rightarrow \gamma \gamma$, but they are not included here as they do not carry any of the conserved charges of strong interactions. An additional note is necessary for $K^0_S$: although the measurement of $K^0_S$ is extremely common in experiments, it is not of use for the treatment we carry on in this work. This is because, from $K^0_S$ only, it is not possible to construct a net-particle quantity (it is its own antiparticle), and additionally part of the information on the mixing between $K^0$ and $\Bar{K^0}$ is lost because $K^0_L$ cannot be measured. For this reason, in the following we will consider $K^0$ and $\Bar{K^0}$ instead, and treat them as ``not-measured''. Finally, we note that, since the decay $\Sigma^0 \rightarrow \Lambda + \gamma$ has a branching ratio of $\sim 100 \%$, effectively what we indicate with $\Lambda$ contains the entire $\Sigma^0$ contribution as well; this well reproduces the experimental situation, where $\Lambda$ and $\Sigma^0$ are treated as the same state. 

It is straightforward to adapt the HRG model so that it is expressed in terms
of stable hadronic states only. The sum over the whole hadronic spectrum is
converted into a sum over both the whole hadronic spectrum, and the list of
states which are stable under strong interactions:
\begin{equation}
\sum_R B_R^l Q_R^m S_R^n I^R_p \rightarrow \sum_{i \in \rm stable} \sum_R \left( P_{R \rightarrow i} \right)^p B^l_i Q^m_i S^n_i I^R_p\, \, ,
\label{eq:daughters}
\end{equation}
with $l+m+n=p$, and where the first sum only runs over the particles which are stable under strong
interactions, and the sum $P_{R \rightarrow i} = \sum_\alpha {\rm N}^\alpha_{R
\rightarrow i} n_{i,\alpha}^R$ gives the average number of particle $i$
produced by each particle $R$ after the whole decay chain. The sum runs over
particle $R$ decay modes, where ${\rm N}^\alpha_{R \rightarrow i}$ is the
branching ratio of the mode $\alpha$, and $n_{i,\alpha}^R$ is the number of
particles $i$ produced by a particle $R$ in the channel $\alpha$.

In light of the above considerations, it is useful to define the contribution  to the conserved charges from final state stable hadrons. In the following, we will adopt the convention where the net-number of particles of species $A$ (i.e., the number of particles $A$ \textit{minus} the number of antiparticles $\overline{A}$) is $\widetilde{A} = A - \overline{A}$.

\begin{figure}
\includegraphics[width=0.99\linewidth]{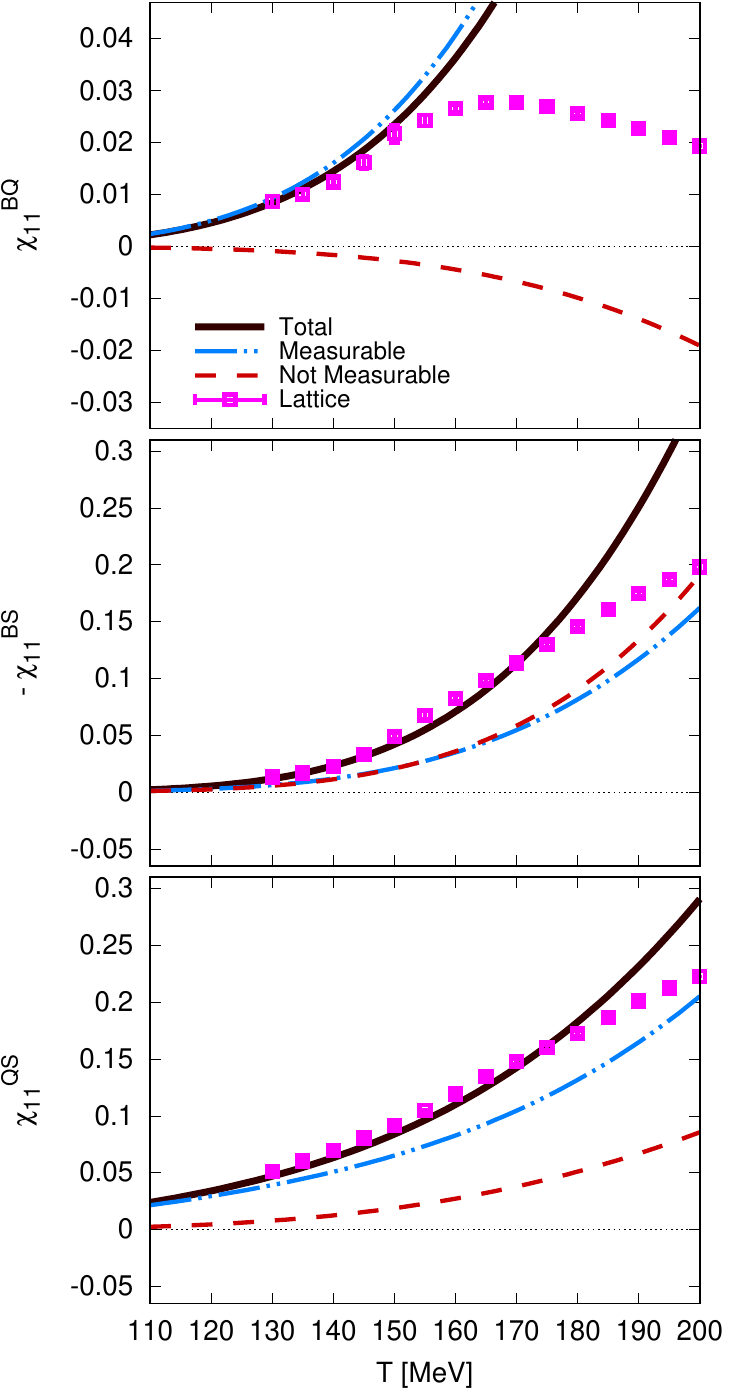}
\caption{Second order correlators of the conserved charges $B,~Q,~S$. The total contribution, the measured and non-measured parts, evaluated in the HRG model, are shown in solid black, dotted-dashed blue and dashed red respectively. The lattice results are shown as the magenta points. }\label{fig:nondiag_corr_nocuts}
\end{figure}

With this definition, we can express conserved charges as:
\begin{align} \label{eq:net-charges}
\text{net-}B:& \quad \widetilde{p} + \widetilde{n} + \widetilde{\Lambda} + \widetilde{\Sigma}^+ + \widetilde{\Sigma}^- + \widetilde{\Xi}^0 + \widetilde{\Xi}^- + \widetilde{\Omega}^-  \, \, , \\ \nonumber
\text{net-}Q:& \quad \widetilde{\pi}^+ + \widetilde{K}^+ + \widetilde{p} + \widetilde{\Sigma}^+ - \widetilde{\Sigma}^- - \widetilde{\Xi}^- - \widetilde{\Omega}^-  \, \, ,  \\ \nonumber
\text{net-}S:& \quad  \widetilde{K}^+ + \widetilde{K}^0 - \widetilde{\Lambda} - \widetilde{\Sigma}^+ - \widetilde{\Sigma}^- - 2 \widetilde{\Xi}^0 - 2 \widetilde{\Xi}^- - 3 \widetilde{\Omega}^-  \, \, .
\end{align}

Using this decomposition, we can write as an example the $BQ$ correlator:
\begin{align} \nonumber \label{eq:corrBQ_def}
\chi^{BQ}_{11} \left(T, \hat{\mu}_B, \hat{\mu}_Q, \hat{\mu}_S \right) &= \sum_R (P_{R \rightarrow \text{net-}B}) (P_{R \rightarrow \text{net-}Q}) \times \\ 
& \qquad \qquad \times I^R_2 \left(T, \hat{\mu}_B, \hat{\mu}_Q, \hat{\mu}_S \right) \, \, ,
\end{align}
where $P_{R \rightarrow \text{net-}B} =  P_{R \rightarrow \widetilde{p}} + P_{R \rightarrow \widetilde{n}} + P_{R \rightarrow \widetilde{\Lambda}} + P_{R \rightarrow \widetilde{\Sigma}^+} + P_{R \rightarrow \widetilde{\Sigma}^-} + P_{R \rightarrow \widetilde{\Xi}^0} + P_{R \rightarrow \widetilde{\Xi}^-} + P_{R \rightarrow \widetilde{\Omega}^-}$, and e.g. $P_{R \rightarrow \widetilde{p}} = P_{R \rightarrow p} - P_{R \rightarrow \overline{p}}$. Analogous expressions apply to $\text{net-}Q$ and $\text{net-}S$.  

\begin{figure}
\includegraphics[width=0.99\linewidth]{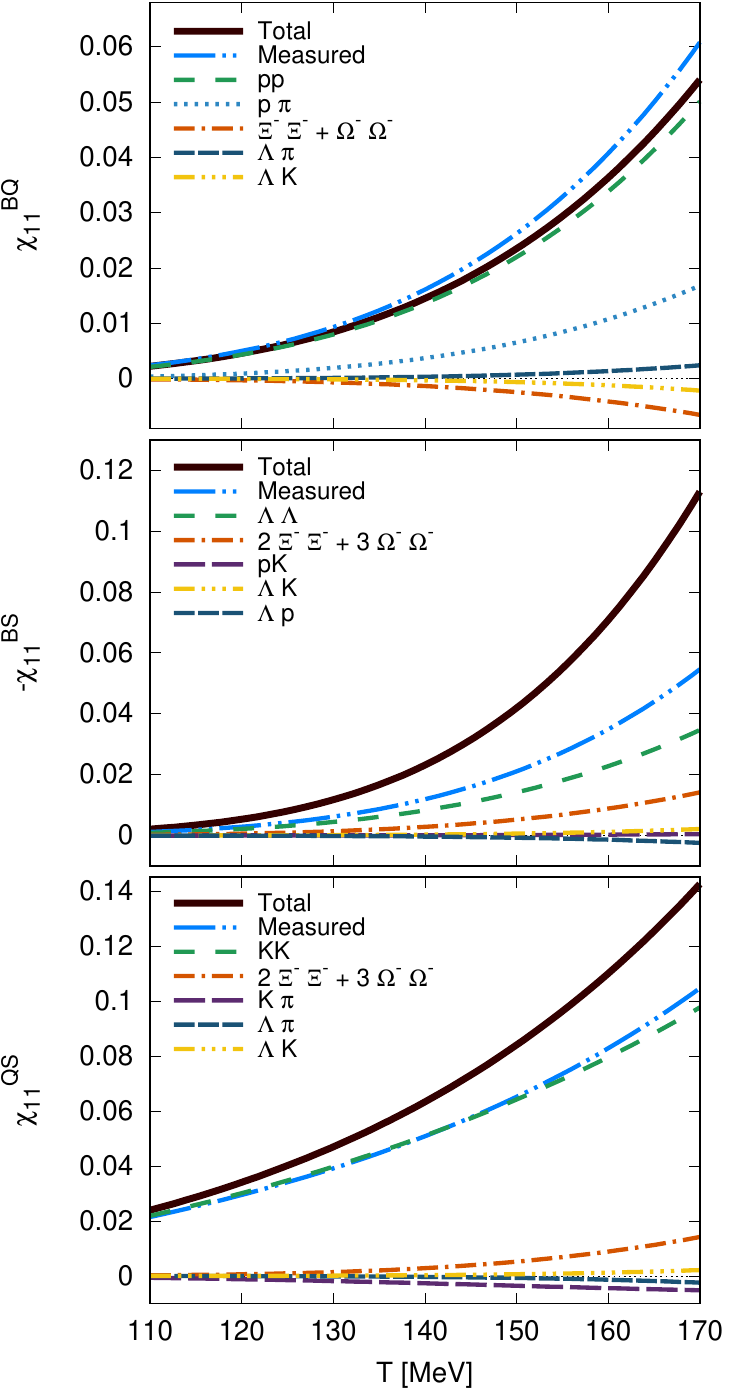}
\caption{Breakdown of the different final state hadronic contributions to the cross correlators of the conserved charges $B,~Q,~S$ at second order. The total contribution and the measured part are shown as solid black and dashed-dotted blue lines respectively. The main single contributions from measured hadronic observables are shown with different colored dashed and dashed-dotted lines. }\label{fig:nondiag_corr_nocuts_brkwdn}
\end{figure}

\begin{figure}
\includegraphics[width=0.99\linewidth]{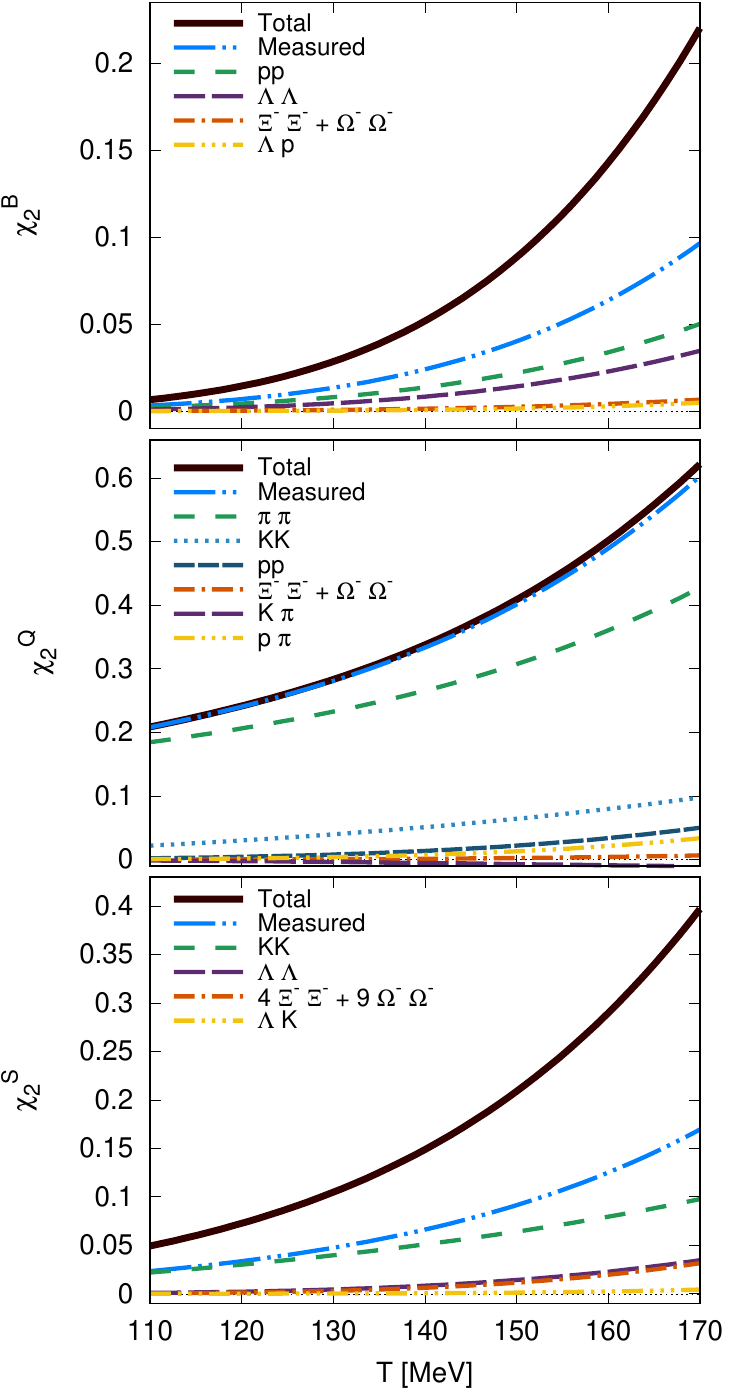}
\caption{Breakdown of the different final state hadronic contributions to the diagonal correlators of the conserved charges $B,~Q,~S$ at second order. The total contribution and the measured part are shown as solid black and dashed-dotted blue lines respectively. The main single contributions from measured hadronic observables are shown with different colored dashed and dashed-dotted lines.}\label{fig:diag_corr_nocuts_brkwdn}
\end{figure}

The result of this decomposition is that each of the correlators one can build between conserved charges, will be formed from the sum of many different particle-particle correlations. In particular, the sum of those correlators which entirely consist of observable species, will yield the ``measured'' part of a certain correlator, while its ``non-measured'' part will consist of all other terms, which include at least one non-observable species.  In Fig. \ref{fig:nondiag_corr_nocuts} the non diagonal correlators are shown as a function of the temperature at vanishing chemical potential. The measured and non-measured contributions are shown with blue, dashed-dotted and red, dashed lines respectively, while the full contribution is shown with a solid, thicker black line.  Alongside the HRG model results, continuum extrapolated lattice results are shown as magenta points as introduced in Section \ref{sec:lattice}.

We notice that both the $BQ$ and $QS$ correlators are largely reproduced by the ``measured'' contribution (for the $BQ$ correlator, the measured portion even exceeds the full one, as the non-measured contribution is negative), while the $BS$ correlator is roughly split in half between measured and non-measured terms. This is, because the former are unsurprisingly dominated by the net-proton and net-kaon contributions respectively, which in this temperature regime form the bulk of particle production, together with the pions. The $BS$ correlator, will conversely receive its main contributions from strange baryons, which are almost equally split between measured and non-measured.

\subsection{Breakdown of the measured and non-measured contributions \label{sec:breakdown_measured}}
The decomposition in Eq. (\ref{eq:net-charges}) allows one to break down the different contributions to any cross correlator, as well as the diagonal ones, entirely. In Figs. \ref{fig:nondiag_corr_nocuts_brkwdn} and \ref{fig:diag_corr_nocuts_brkwdn}, we show the breakdown of the measured portion of the single final state hadronic (self) correlations to the non-diagonal and diagonal correlators respectively. Let us start from the non-diagonal case. 

A few features can be readily noticed. First, in all cases only a handful of the most sizable contributions account for the measured portion of the corresponding observable. As stated above, the $BQ$ and $QS$ correlators are expected to be dominated by the contribution from net-proton and net-kaon self-correlations respectively: indeed, in both cases the measured part almost entirely consists of these major contributions. Second, it is worth noticing how, with the only exception of the proton-pion correlator within $\chi_{11}^{BQ}$, all correlators between different species yield a very modest contribution. This is the case for the proton-kaon, kaon-pion, Lambda-pion and Lambda-kaon correlators in $\chi_{11}^{BQ}$ and $\chi_{11}^{QS}$, as well as theproton-kaon, Lambda-kaon and Lambda-proton correlators in $\chi_{11}^{BS}$.. In our setup, correlations between different particle species can only arise from the decay of heavier resonances. Whenever a resonance $R$ has a non-zero probability to decay, after the whole decay cascade, both into stable species $A$ and $B$, then a correlation arises between $A$ and $B$. It can be seen from Eq. (\ref{eq:corrBQ_def}) that only when both probabilities in parentheses are non-zero, a non-zero correlation can arise. For the same reason, correlations between different baryons arise, although no single decay mode with more that one baryon (or anti-baryon) is present in our decay list. In fact, if a state exists which has a finite probability to produce -- after the whole decay cascade -- both baryon $A$ and baryon $B$, then a correlation between $A$ and $B$ is generated through Eq. (\ref{eq:corrBQ_def}). Finally, since both $\Xi^-$ and $\Omega^-$ carry all three conserved charges, they contribute to all three correlators through their self-correlations, and their contribution is not negligible in all cases. 

The case of $\chi_{11}^{BS}$ is slightly different, as the measured part is smaller than in the cases of $\chi^{QS}_{11}$ and $\chi^{BQ}_{11}$. This is due to the fact that there is no significant separation in mass between the lightest observable particle carrying both baryon number and strangeness -- the $\Lambda$ baryon -- and the lightest of the non-measured ones -- the $\Sigma^\pm$ baryons. In fact, the contribution from both charged $\Sigma$ baryons is comparable to the one from the $\Lambda$, which thus cannot play as big of a role as the proton and kaon in the other two correlators, as well as because of the previously mentioned fact that correlators of different species do not contribute significantly. 

In the diagonal case, a similar picture appears. The $\chi_2^Q$ correlator is almost identical to its measured portion, dominated by the self correlations of pions, kaons, and protons.  The other two correlators have a similar situation to that of $\chi_{11}^{BS}$, with the measured part roughly amounting to half of the total. Again, the only non-negligible correlator between different species is the proton-pion correlator in $\chi_2^{Q}$. We notice that in general, the leading single contribution is not as close to the whole measured portion, as it was in the case of the cross correlators. This aspect will be important in the following, where we will move to the analysis of ratios of correlators, and look for suitable proxies.

\begin{figure}
\includegraphics[width=0.99\linewidth]{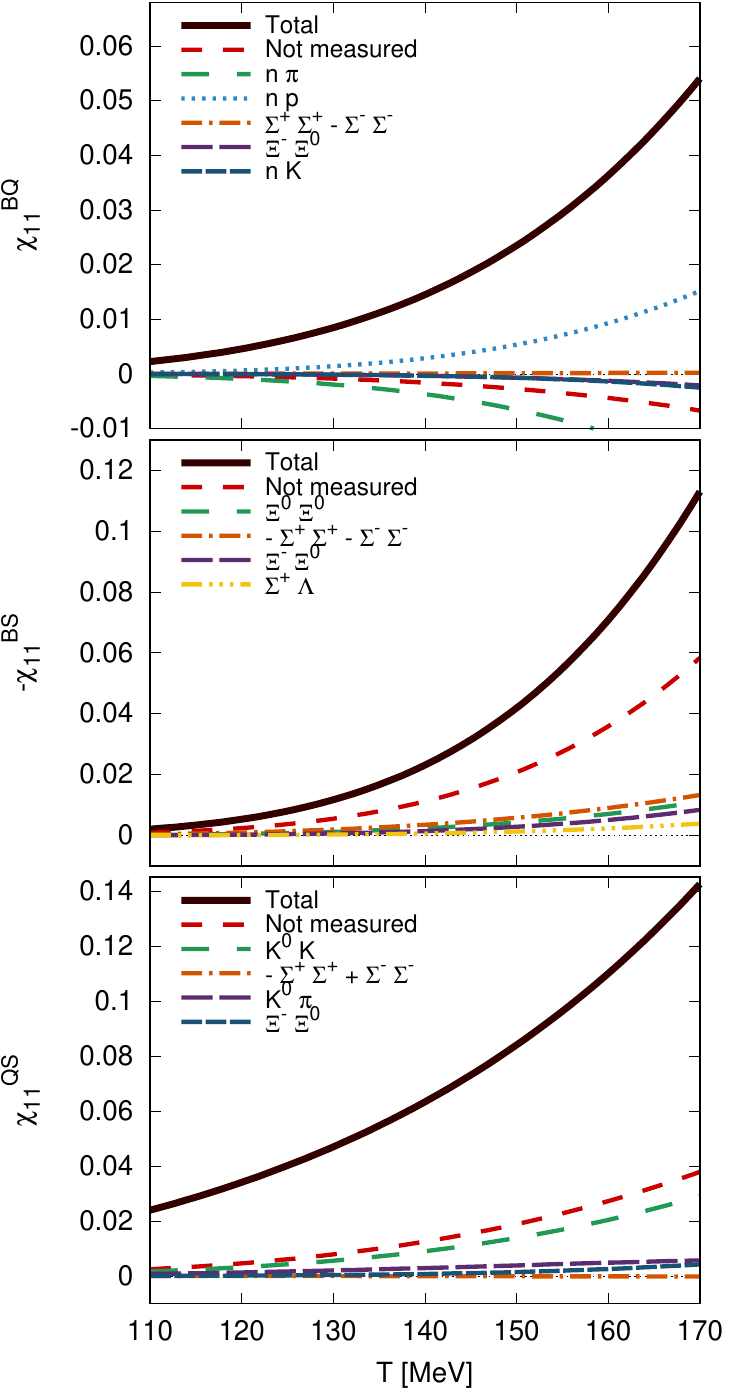}
\caption{Breakdown of the different final state hadronic contributions to the cross correlators of the conserved charges $B,~Q,~S$ at second order. The total contribution and the non-measured part are shown as solid black and dashed red lines respectively. The main single contributions from non-measured hadronic observables are shown with different colored dashed and dashed-dotted lines. }\label{fig:nondiag_corr_nocuts_brkwdn_nomeas}
\end{figure}

\begin{figure}
\includegraphics[width=0.99\linewidth]{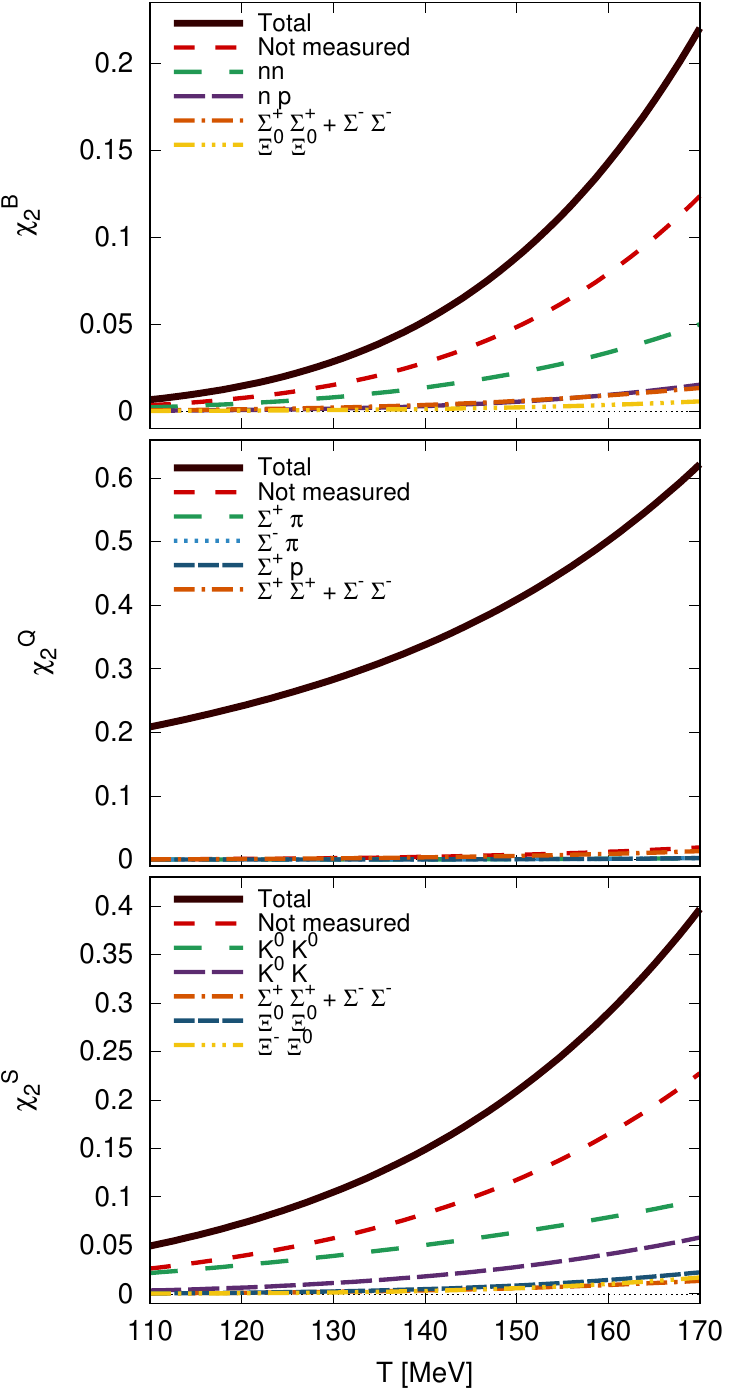}
\caption{Breakdown of the different final state hadronic contributions to the diagonal correlators of the conserved charges $B,~Q,~S$ at second order. The total contribution and the non-measured part are shown as solid black and dashed red lines respectively. The main single contributions from non-measured hadronic observables are shown with different coloured dashed and dashed-dotted lines. }\label{fig:diag_corr_nocuts_brkwdn_nomeas}
\end{figure}

In Figs. \ref{fig:nondiag_corr_nocuts_brkwdn_nomeas} and
\ref{fig:diag_corr_nocuts_brkwdn_nomeas} we show the breakdown of the
non-measured portion of the final state hadronic (self) correlations,
analogously to what we showed in Figs. \ref{fig:nondiag_corr_nocuts_brkwdn} and
\ref{fig:diag_corr_nocuts_brkwdn} for the measured portion. The
situation in this case is slightly different from the previous one: it is
generally more difficult to identify a leading contribution, with multiple
terms yielding comparable results. In the case of $BQ$ and $QS$, leading terms
come with opposite signs, which further complicates the picture. The reasons for 
these features come from the 
fact that: i) the number of single contributions that are not measured is much larger 
than that of the measured ones, hence it is less probable that few terms dramatically
dominate; ii) in general, non-measured species are heavier than the measured ones, 
hence single contributions tend to be smaller. Obviously, exceptions to this are the 
neutron and $K^0$. In fact, the diagonal correlators $\chi_2^B$ and $\chi_2^S$ show a 
sizable input from $\sigma_n^2$ (the variance of the neutron distribution), and 
both $\sigma_{K^0}^2$ and
$\sigma_{K^0 K}$, respectively. The case of $\chi_2^Q$ is peculiar since, as
evident from Fig. \ref{fig:nondiag_corr_nocuts}, the non-measured
contribution is almost negligible, when compared to the measured one.

\subsection{\label{sec:randomization}Isospin randomization}

Another important effect we have not addressed yet, which is present in experiment, is the isospin randomization \cite{Kitazawa:2011wh,Kitazawa:2012at}. This effect is caused by reactions that take place in the hadronic phase between nucleons and pions, and consist of the generation and decay of $\Delta$ resonances ($\Delta (1232)$ prominently), through processes like:
\begin{align} \label{eq:randomization} \nonumber
p + \pi^0 &\leftrightarrow \Delta^+ \leftrightarrow n + \pi^+\,\,,\\
p + \pi^- &\leftrightarrow \Delta^0 \leftrightarrow n + \pi^0\,\,,
\end{align}
with analogous ones for the anti-baryons. For collision energies $\sqrt{s}
\gtrsim 10 \GeV$, the lifetime of the fireball is long enough to allow several
of such cycles to take place, resulting in a complete randomization of the isospin
of the nucleons \cite{Kitazawa:2011wh,Kitazawa:2012at}.  
This expectation has been confronted with data in \cite{Nahrgang:2014fza}
confirming a complete randomization with the exception of the highest
energy: $\sqrt{s}=200~\mathrm{GeV}$. For this paper, though, we will assume 
complete randomization throughout. 

The distributions of protons and neutrons then factorize and the correlation
between the two is erased. The average number of protons and neutrons, as well
as anti-protons and anti-neutrons, and consequently the average net-proton and
net-neutron number, are left unchanged by such reactions, but fluctuations are
not.  In particular, this results in an enhancement of both the net-proton and
net-neutron variance, at the expense of the correlation between the two (note
that the variance of net-nucleon $\sigma^2_N = \sigma^2_p + 2 \sigma_{pn} +
\sigma^2_n$ cannot be changed by these reactions). Similarly, charge
conservation ensures that the sum $\widetilde{Q} = \widetilde{p} +
\widetilde{\pi}$ is conserved in the reactions in Eq. (\ref{eq:randomization}).
It can be shown that this results in the net-pion variance $\sigma^2_\pi$ being 
increased by the same amount as the net-proton variance$\sigma^2_p$. 
Since the sum $\sigma^2_{p+\pi} = \sigma^2_p + 2
\sigma_{p\pi} + \sigma^2_\pi$ must also be left unchanged, we have that
$\sigma_{p\pi}$ is decreased by the same amount again: $\sigma_{pn}$.

Thus there is information lost through the process of randomization, and
the original $\sigma_{pn}$ or $\sigma_{p}^2$ cannot be reproduced
from data individually. The experimental access to those correlators where either of these plays an important role is very difficult. This is the case e.g. for $\chi_{11}^{BQ}$, which is completely dominated by $\sigma_p^2$.

\section{Proxies} \label{sec:proxies}

The issue of having particles that cannot be detected poses the problem of a loss of conserved charges. Historically, the proxies for baryon number, electric charge, and strangeness have been the protons, the $p,\pi,K$ combination, and the kaons themselves respectively. 

\begin{figure}
\center
\includegraphics[width=\linewidth]{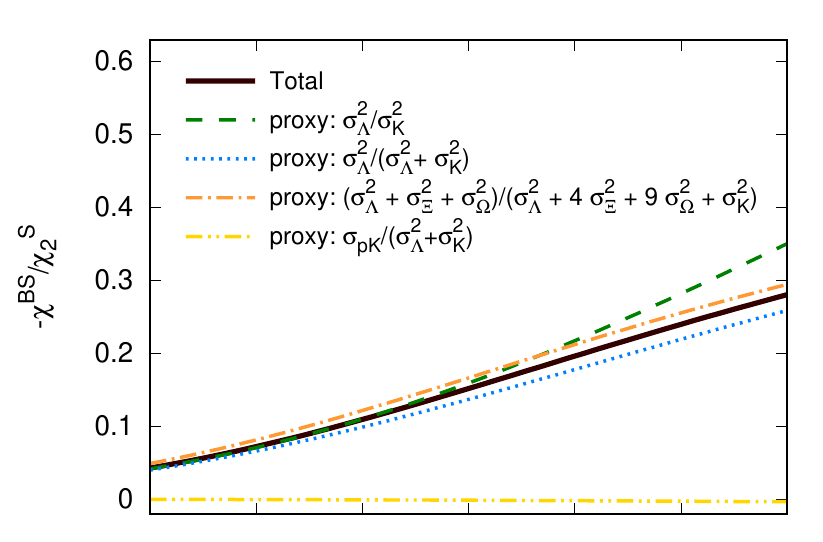} \\ \vspace{-5mm}
\includegraphics[width=\linewidth]{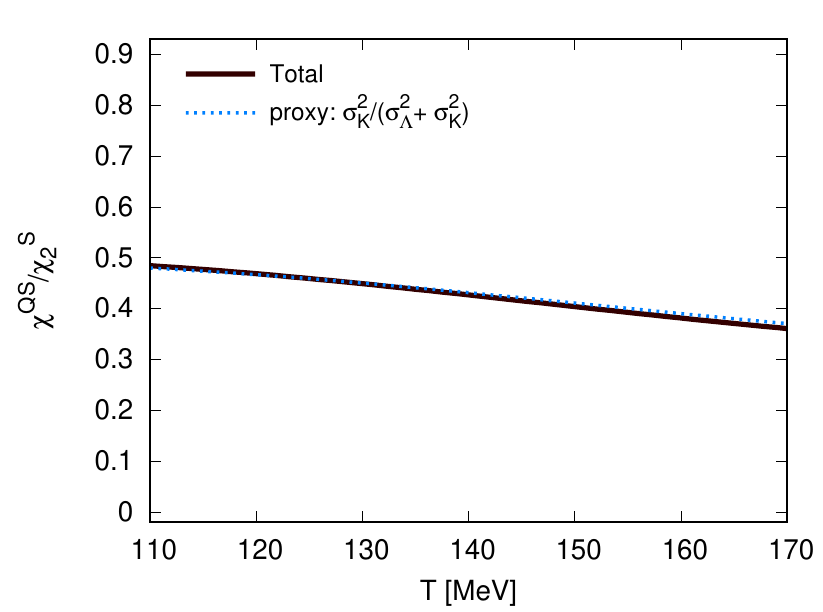}
\caption{The temperature dependence of the ratios $-\chi_{11}^{BS}/\chi_2^S$ (upper panel) and $\chi_{11}^{QS}/\chi_2^S$ (lower panel), at $\mu_B = 0$. In both cases the total contribution is shown with a solid black line, along with different proxies: (upper panel) $\widetilde{C}^{\Lambda,K}_{BS,SS}$ (dashed green line), $\widetilde{C}^{\Lambda,\Lambda K}_{BS,SS}$ (dotted blue line), $\widetilde{C}^{\Lambda\Xi\Omega, \Lambda\Xi\Omega K}_{BS,SS}$ (dashed-dotted orange line) and $\widetilde{C}^{pK,\Lambda K}_{BS,SS}$ (dashed-double-dotted yellow line), defined in Eq.s (\ref{eq:pxy_BS_SS_1}), (\ref{eq:pxy_BS_SS_2}), (\ref{eq:pxy_BS_SS_3}), and (\ref{eq:pxy_BS_SS_4}), respectively; (lower panel) $\widetilde{C}^{K,\Lambda K}_{QS,SS}$ (dotted blue line) defined in Eq. (\ref{eq:pxy_QS_SS_1}).} \label{fig:BQS_ratios_pxy}
\end{figure}

We have seen in Figs. \ref{fig:nondiag_corr_nocuts_brkwdn} and \ref{fig:diag_corr_nocuts_brkwdn} how the single hadronic, measured (self) correlators relate to the fluctuations of conserved charges. We can then find a correspondence between fluctuations of conserved charges and measurable (and calculable) hadronic fluctuations. 

Both in theory and experiment, it is customary to consider ratios of fluctuations, in order to eliminate, at least at leading order, the dependence on the system volume. For this reason, we will focus on the ratios $\chi_{11}^{BS}/\chi_2^S$ and $\chi_{11}^{QS}/\chi_2^S$, for which we would like to construct proxies using solely fluctuations of (measured) hadrons. The underlying assumption when considering ratios is that the freeze-out of all species involved occurs at the same time in the evolution of the system, hence at the same volume.

Let us start considering the $\chi_{11}^{BS}$ correlator. One could expect that, having both kaons and protons in the bulk of particle production, their correlator $\sigma_{pK}$ would be a good proxy. However, as we can see in Fig. \ref{fig:nondiag_corr_nocuts_brkwdn}, this is clearly not the case, as the proton-kaon correlator gives a negligible contribution to $\chi_{11}^{BS}$. On the contrary, the variance of the net-Lambda distribution $\sigma_\Lambda^2$ represents a much more sizable contribution to the total correlator. 

In the upper panel of Fig. \ref{fig:BQS_ratios_pxy} we show the HRG model results for the ratio $\chi_{11}^{BS}/\chi_2^S$ at $\mu_B = 0$ (black, thicker line). As already mentioned, from Figs. \ref{fig:nondiag_corr_nocuts_brkwdn} and \ref{fig:diag_corr_nocuts_brkwdn} we see how the leading contributions to the two correlators come from $\sigma_\Lambda^2$ and $\sigma_K^2$ respectively. We can then construct a tentative proxy as:
\begin{equation} \label{eq:pxy_BS_SS_1}
\widetilde{C}^{\Lambda,K}_{BS,SS} = \sigma_\Lambda^2/\sigma_K^2 \, \, ,
\end{equation} 
which is shown as a green, dashed line. We see that, although this quantity reproduces very well the full result at low temperatures -- where the kaons dominate -- it overshoots at higher temperatures, and in particular around the QCD transition and chemical freeze-out temperatures, which are obviously the interesting regime. It is worth noticing that, in order to construct a good proxy for a ratio of conserved charges fluctuations, it is not sufficient to choose the best proxy for both the numerator and the denominator. In fact, a good proxy for the ratio will be obtained when the proxy in the numerator and the denominator are equally good. Some guidance in this construction is then provided by Fig. \ref{fig:diag_corr_nocuts_brkwdn}, where the extent to which a hadronic correlator reproduces the corresponding $BQS$ fluctuation is most evident. For this reason, we consider adding the contribution from the net-$\Lambda$ fluctuations to $\chi_2^S$ too, and define:
\begin{equation}\label{eq:pxy_BS_SS_2}
\widetilde{C}^{\Lambda,\Lambda K}_{BS,SS} = \sigma_\Lambda^2/( \sigma_K^2 + \sigma_\Lambda^2) \, \, ,
\end{equation}
which is shown as a blue, dotted line. We see how this second proxy is much better at reproducing the full result, as it is very close to it at all temperatures, including in the vicinity of the QCD transition. In addition, again referring to Figs. \ref{fig:nondiag_corr_nocuts_brkwdn} and \ref{fig:diag_corr_nocuts_brkwdn}, it is interesting to try and include the contributions from multi-strange hadrons, both in the numerator and denominator. With these, one has:
\begin{equation} \label{eq:pxy_BS_SS_3}
\widetilde{C}^{\Lambda\Xi\Omega, \Lambda\Xi\Omega K}_{BS,SS} = (\sigma_\Lambda^2 + 2 \sigma_\Xi^2 + 3 \sigma_\Omega^2)/(\sigma_\Lambda^2 + 4 \sigma_\Xi^2 + 9 \sigma_\Omega^2 + \sigma_K^2) \, \, ,
\end{equation}
which is shown as the orange, dashed-dotted line, and also reproduces very well the behavior of the full ratio, although not really improving the situation over the previous one. As a final check, one can build a proxy from the $\sigma_{pK}$ correlator as:
\begin{equation} \label{eq:pxy_BS_SS_4}
\widetilde{C}^{pK,\Lambda K}_{BS,SS} = \sigma_{11}^{pK}/( \sigma_K^2 + \sigma_\Lambda^2) \, \, ,
\end{equation}
which is shown as the yellow, dashed-double-dotted line. Not unexpectedly, this combination is not able to serve as a good proxy. 

The case of $\chi_{11}^{QS}/\chi_2^Q$ follows directly from the previous one, and is shown in the lower panel of Fig. \ref{fig:BQS_ratios_pxy}. In fact, in a system with $2+1$ quarks (with no isospin symmetry breaking) the following relation applies:
\begin{equation}
2 \left\langle Q S \right\rangle - \left\langle BS \right\rangle = \left\langle SS \right\rangle \, \, ,
\end{equation}
from which one can derive that:
\begin{equation}
\frac{\chi_{11}^{QS}}{\chi_2^S} = \frac{1}{2} \left( 1 - \frac{\chi_{11}^{BS}}{\chi_2^S} \right) \, \, .
\end{equation}
Thus, exploiting this relation and the good proxy $\widetilde{C}^{\Lambda,\Lambda K}_{BS,SS}$ we have defined for $\chi_{11}^{BS}/\chi_2^S$, we can define:
\begin{equation} \label{eq:pxy_QS_SS_1}
\widetilde{C}^{K, \Lambda K}_{QS,SS} = \frac{1}{2} \sigma_K^2/(\sigma_\Lambda^2 + \sigma_K^2) \, \, ,
\end{equation}
and have a proxy for $\chi_{11}^{QS}/\chi_2^S$ for free, which indeed works very well over the whole temperature range.

\begin{figure}
\includegraphics[width=\linewidth]{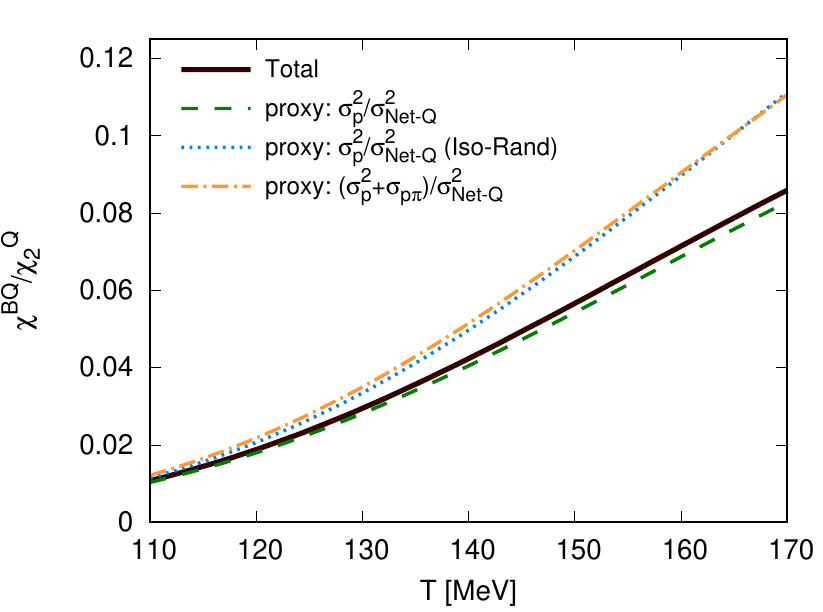}
\caption{Temperature dependence of the ratio $\chi_{11}^{BQ}/\chi_2^Q$ at $\mu_B=0$ (solid black line), together with the ratios $\widetilde{C}^{p,\text{Net-}Q}_{BQ,QQ}$, both with (dashed green line) and without (dotted blue line) isospin randomization, and $\widetilde{C}^{p \pi,\text{Net-}Q}_{BQ,QQ}$ (dotted blue line), defined in Eqs. (\ref{eq:pxy_BQ_QQ}) and (\ref{eq:pxy_BQ_QQ_2}) respectively.}
\label{fig:BQ_QQ_pxy}
\end{figure}

Now, we wish to consider the correlator $\chi_{BQ}$, which is the only one of the three non-diagonal correlators to be influenced by the isospin randomization discussed in the previous Section. By looking at Figs. \ref{fig:nondiag_corr_nocuts_brkwdn} and \ref{fig:diag_corr_nocuts_brkwdn}, it is natural to construct the proxy
\begin{equation}  \label{eq:pxy_BQ_QQ}
\widetilde{C}^{p, \rm{Net}-Q}_{BQ,QQ} = \sigma_p^2/ \sigma^2_{\rm Net-Q}\, \, ,
\end{equation}
where the net-charge is typically defined as $\widetilde{Q} = \widetilde{p} + \widetilde{\pi} + \widetilde{K}$. In Fig. \ref{fig:BQ_QQ_pxy} the total contribution and this proxy are shown with solid, thick black line and a dashed green line, respectively, and the agreement is extremely good. When including the effect of isospin randomization -- which does not affect the denominator -- the situation is radically different, and the corresponding curve is shown with a dotted blue line. The increase in the net-proton variance spoils the effectiveness of this proxy. A quantity which is not affected by this effect is the sum $\sigma^2_p + \sigma_{p\pi}$, from which we can define the ratio:
\begin{equation}  \label{eq:pxy_BQ_QQ_2}
\widetilde{C}^{p\pi, \rm{Net}-Q}_{BQ,QQ} = (\sigma_p^2 + \sigma_{p\pi})/ \sigma^2_{\rm Net-Q}\, \, .
\end{equation}
This quantity is also shown in Fig. \ref{fig:BQ_QQ_pxy} as a dashed-dotted
orange line, and clearly cannot serve as a good proxy. It is interesting to
notice how the increase that $\sigma^2_p$ receives by the isospin randomization
is almost exactly equal to $\sigma_{p\pi}$. As we discussed in Section \ref{sec:randomization} the effect of isospin randomization on $\sigma^2_p$ amounts to 
$\sigma_{pn}$. The presently used HRG-based approach introduces a $p-\pi$ ($\sigma_{p\pi}$) and $p-n$ ($\sigma_{pn}$) correlation through the decay of the same resonances ($\Delta$).
 
From Fig. \ref{fig:BQ_QQ_pxy} we see that, because of this effect, it is not possible to build a suitable proxy for $\chi_{11}^{BQ}/\chi_2^Q$. For analogous reasons, it is not possible to create a good proxy for the ratio $\chi_{11}^{BQ}/\chi_2^B$.

\begin{figure}
\includegraphics[width=\linewidth]{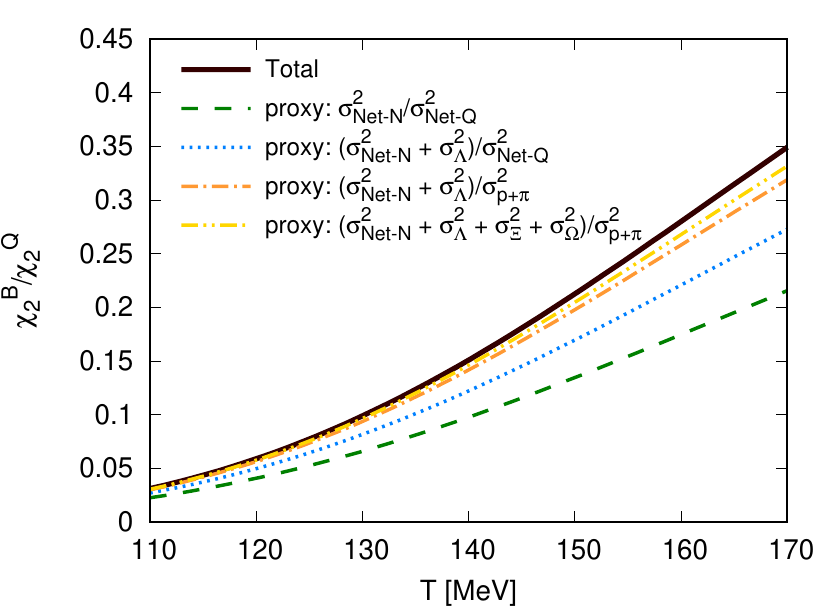}
\caption{Temperature dependence of the ratio $\chi_2^B/\chi_2^Q$ at $\mu_B=0$ (solid black line), together with the ratios $\widetilde{C}^{\rm{Net}-N, \rm{Net}-Q}_{BB,QQ}$ (dashed green line), $\widetilde{C}^{\rm{Net}-N \Lambda, \rm{Net}-Q}_{BB,QQ}$ (dotted blue line), $\widetilde{C}^{\rm{Net}-N \Lambda, p\pi}_{BQ,QQ}$ (dashed-dotted orange line), and $\widetilde{C}^{\rm{Net}-N \Lambda \Xi \Omega, p\pi}_{BQ,QQ}$ (dashed-double-dotted yellow line) defined in Eqs. (\ref{eq:pxy_BB_QQ}), (\ref{eq:pxy_BB_QQ_2}), (\ref{eq:pxy_BB_QQ_3}) and (\ref{eq:pxy_BB_QQ_4}) respectively.}
\label{fig:BB_QQ_pxy}
\end{figure}

Having discussed all three combinations of the off-diagonal cross-correlators
we are lacking a good proxy for a correlator ratio involving only the light
quarks. As a detour from the main line of the discussion we show that this
is also a difficult task in the case of the diagonal correlators.
Consider the ratio $\chi_2^B/\chi_2^Q$. In Fig. \ref{fig:BB_QQ_pxy} we see the
temperature dependence of this ratio at $\mu_B=0$, and the behavior of some
tentative proxies alongside it. We start by considering the quantity:
\begin{equation}  \label{eq:pxy_BB_QQ}
\widetilde{C}^{\rm{Net}-N, \rm{Net}-Q}_{BB,QQ} = \sigma^2_{\rm{Net}-N}/ \sigma^2_{\rm Net-Q}\, \, .
\end{equation}
where we take advantage of the fact that, after the isospin randomization, one has $\sigma^2_{\rm {Net}-N} = 2 \sigma^2_p$. This quantity is shown in Fig. \ref{fig:BB_QQ_pxy} as a green dashed line, and we see that its contribution is not sufficient. We then consider adding the contribution from $\Lambda$ baryons, and show as a dotted blue line the quantity:
\begin{equation}  \label{eq:pxy_BB_QQ_2}
\widetilde{C}^{\rm{Net}-N \Lambda, \rm{Net}-Q}_{BB,QQ} = (\sigma^2_{\rm{Net}-N} + \sigma^2_\Lambda)/ \sigma^2_{\rm Net-Q}\, \, ,
\end{equation}
which improves on the previous one, but is still not satisfactory. We then try removing the contribution from net-kaons at the denominator -- which we can do regardless of isospin randomization:
\begin{equation}  \label{eq:pxy_BB_QQ_3}
\widetilde{C}^{\rm{Net}-N \Lambda, p\pi}_{BB,QQ} = (\sigma^2_{\rm{Net}-N} + \sigma^2_\Lambda)/( \sigma^2_p + 2 \sigma_{p\pi} + \sigma^2_\pi)\, \, ,
\end{equation}
and finally include the contribution from multi-strange baryons in the numerator:
\begin{equation}  \label{eq:pxy_BB_QQ_4}
\widetilde{C}^{\rm{Net}-N \Lambda \Xi \Omega, p\pi}_{BB,QQ} = (\sigma^2_{\rm{Net}-N} + \sigma^2_\Lambda + \sigma^2_\Xi + \sigma^2_\Omega)/( \sigma^2_p + 2 \sigma_{p\pi} + \sigma^2_\pi)\, \, .
\end{equation}

These last two proxies are also shown in Fig \ref{fig:BB_QQ_pxy} as an orange
dashed-dotted and as a yellow-double-dotted line respectively. We see that both
compare relatively well with the total contribution, with the latter being the
better one. It is quite interesting how difficult it was to construct a
suitable proxy for light-quark-dominated observables, in comparison to the
previous cases of $\chi_{11}^{BS}$ and $\chi_{11}^{QS}$. This is mainly due to the
fact that i) net-charge is such a good proxy for $\chi_2^Q$ that is hard to match
for other correlators, and ii) isospin randomization prevents from building
proxies with fluctuations of net-proton only.

We have seen in this Section how  to construct good proxies for ratios
including both diagonal and cross correlators of conserved charges. The proxies
including strangeness make use of only a couple of hadronic observables, namely
the variances $\sigma^2_K$ and $\sigma^2_\Lambda$ -- more precisely, only their
ratio.  It is also remarkable how the addition of multi-strange baryons to the
proxy for $\chi_{11}^{BS}/\chi_2^S$ is not necessary, as it does not improve
the existing agreement. We also saw that for light-quark-dominated observables,
isospin randomization modifies the correlators of net-proton, net-pion and
net-neutron, preventing the construction of useful proxies for such
observables. 

\begin{figure*}
\center
\includegraphics[width=\textwidth]{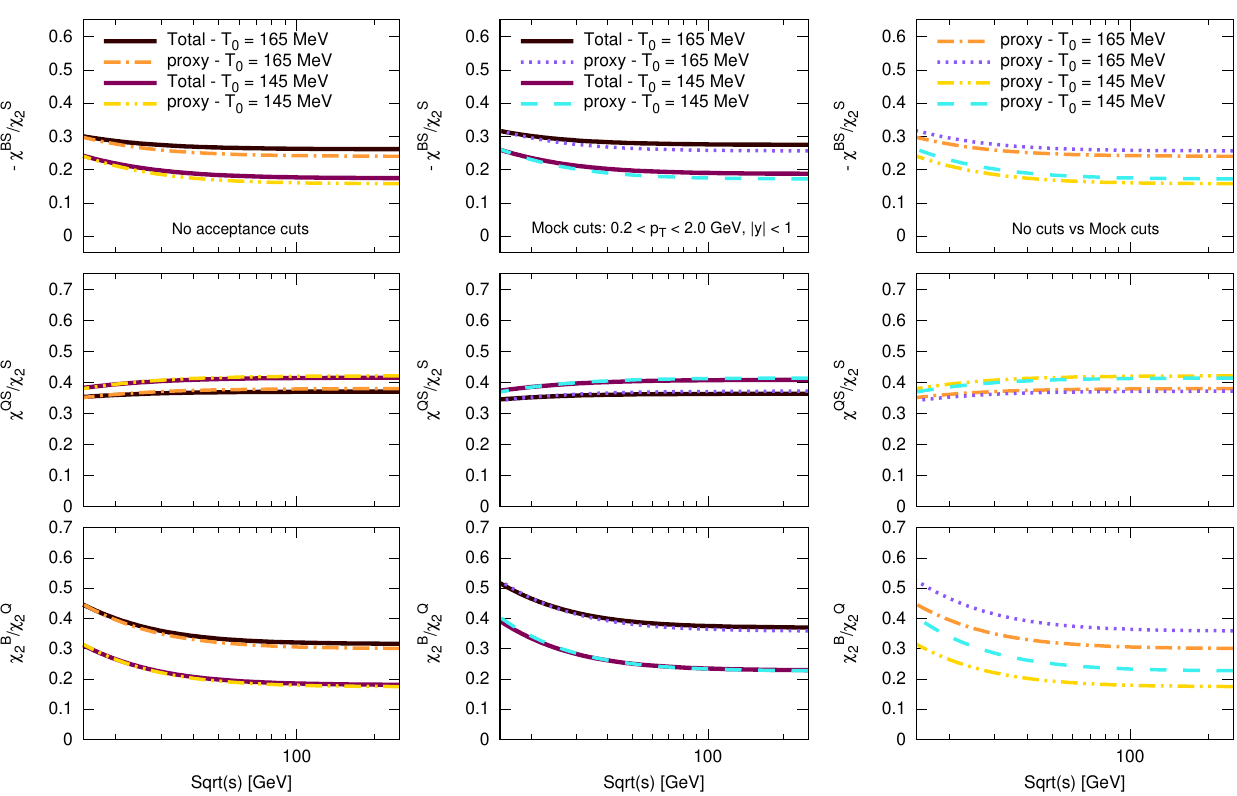}
\caption{Behavior of the ratios $-\chi_{11}^{BS}/\chi_2^S$ and
$\chi_{11}^{QS}/\chi_2^S$ along parametrized chemical freeze-out lines with
$T_0 = 145 \MeV$ and $T_0 = 165 \MeV$.  For a comparison we also show the
diagonal $\chi^{B}_2/\chi^Q_2$ in the third row.  For each ratio, the best proposed proxy is shown as well, for both temperatures:
$\widetilde{C}^{\Lambda,\Lambda K}_{BS,SS}$ for $-\chi_{11}^{BS}/\chi_2^S$,
$\widetilde{C}^{K,\Lambda K}_{QS,QQ}$ for $\chi_{11}^{QS}/\chi_2^S$ and
$\widetilde{C}^{N \Lambda \Xi \Omega,p\pi}_{BB,QQ}$ for $\chi_2^B/\chi_2^Q$.
In the
left panel, we show the results in the case without kinematic cuts: the total
contribution is shown with black and burgundy solid lines for $T_0 = 145
\MeV$ and $T_0 = 165 \MeV$, respectively; the proxy is shown with a yellow
dashed-double-dotted and orange dashed-dotted line for $T_0 = 145 \MeV$ and
$T_0 = 165 \MeV$. In the central panel, we show the results in the case with
the ``mock'' cuts discussed in the text: in this case the proxy is shown with a
cyan dashed and purple dotted line for $T_0 = 145 \MeV$ and $T_0 = 165 \MeV$.
Finally, in the right panel we compare the behavior of the proxies with and
without the introduction of cuts, and keep the same color code as from the
right and central panel.}
\label{fig:BQS_ratios_pxy_FO}
\end{figure*}

\section{Finite chemical potential and kinematic cuts} \label{sec:finite_mu_cuts}

Since experimental measurements for moments of net-particle distributions are
currently available both from the LHC and RHIC, it is interesting to analyze
the behavior of the quantities we are studying also at finite values of the
baryon chemical potential. In the left panel of Fig.
\ref{fig:BQS_ratios_pxy_FO}, we show the behavior of the proxies along
parametrized chemical freeze-out lines -- shifted in $T$ from the
parametrization in \cite{Cleymans:2004pp} -- with $T$ intersects at $T_0 = 145,
165 \MeV$, so to ``bracket'' the crossover region of QCD. The ratios
$\chi_{11}^{BS}/\chi_2^S$ and $\chi_{11}^{QS}/\chi_2^S$
are shown in the first and second row respectively. We see that for
these ratios, the agreement with the considered proxies does not worsen with
the increase in the chemical potential, and the curves remain very close for a
broad range of collision energies. This means that the scope of the proxies we
have constructed to reproduce the behavior of fluctuations of conserved charges
is not limited to small $\mu_B$, but can be extended to the study in the BES as
well. 

In Section \ref{sec:HRG2} we have mentioned that one of the strengths of the
HRG model is the possibility it offers to include effects that are present in
the experimental situation, like the use of cuts on the kinematics. In the
central panel of Fig. \ref{fig:BQS_ratios_pxy_FO} we show the same scenario as
in the left panel, but with the inclusion of exemplary, ``mock cuts'': $0.2
\leq p_T \leq 2.0 \GeV$, $\left| y \right| \leq 1.0$. These cuts do not
correspond to any past or ongoing measurement at LHC or RHIC, but are
constructed such to be reproducible in the experiment, and still give a hint of
the effect of including the cuts at all. For a systematic treatment of the
dependence of fluctuations on the kinematic cuts -- which is beyond the scope
of this work -- see \cite{Chatterjee:2016mve}, where it is studied in a thermal
model with an older hadron list and without the inclusion of resonance decays.
In our example, the same cuts are applied to all particle species. We see that
for all the observables considered the agreement between net-charge fluctuation
ratios and proxies remains the same as in the case without cuts, for both
freeze-out lines. 

Finally, in the right panel of Fig. \ref{fig:BQS_ratios_pxy_FO} we show the
selected proxies, for both freeze-out curves, comparing the cases with and
without the cuts. We see that the effect is very minimal for the two ratios
$\chi^{BS}_{11}/\chi^{S}_2$ and $\chi^{QS}_{11}/\chi^{S}_2$. This is obviously
of key importance in light of a potential direct comparison to results from
lattice QCD calculations, as the one discussed in Section \ref{sec:finitemu}
for $\chi^{BS}_{11}/\chi^{S}_2$. This is one of the main reasons these proxies
were build in the first place.

The third row in Fig.~\ref{fig:BQS_ratios_pxy_FO} shows the behaviour of the
ratio $\chi_2^B/\chi_2^Q$ when acceptance cuts are introduced. As opposed to
the discussed off-diagonal ratios it shows a large dependence on the cuts.
Thus, even though this ratio does not suffer from the effect of isospin
randomization, a comparison to lattice simulations can be problematic.
Thus, we will focus on the strangeness related off-diagonal correlators
in the next section, where we compare to experimental data.

\section{Comparison to experimental results} \label{sec:exp_comp}

In the previous Section we have considered the impact of including kinematic
cuts on the proxies we have defined previously, by considering some exemplary
cuts which were chosen to be the same for all particle species. However,
experimental measurements exist for different species, and it is possible to
test how the proxies we constructed compare to the experimental results, this
time including the corresponding cuts on a species-by-species (or
measurement-by-measurement) basis. 

\begin{figure}
\center
\includegraphics[width=0.95\linewidth]{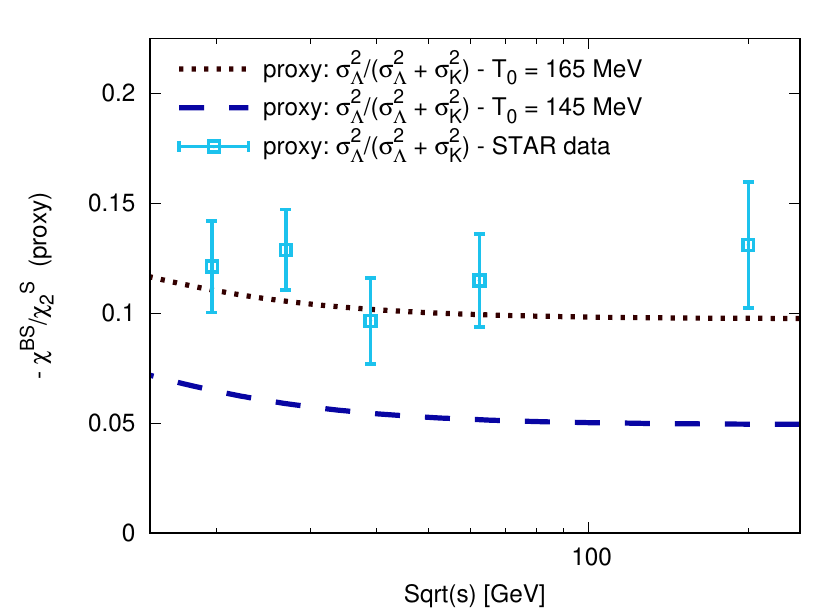}
\includegraphics[width=0.95\linewidth]{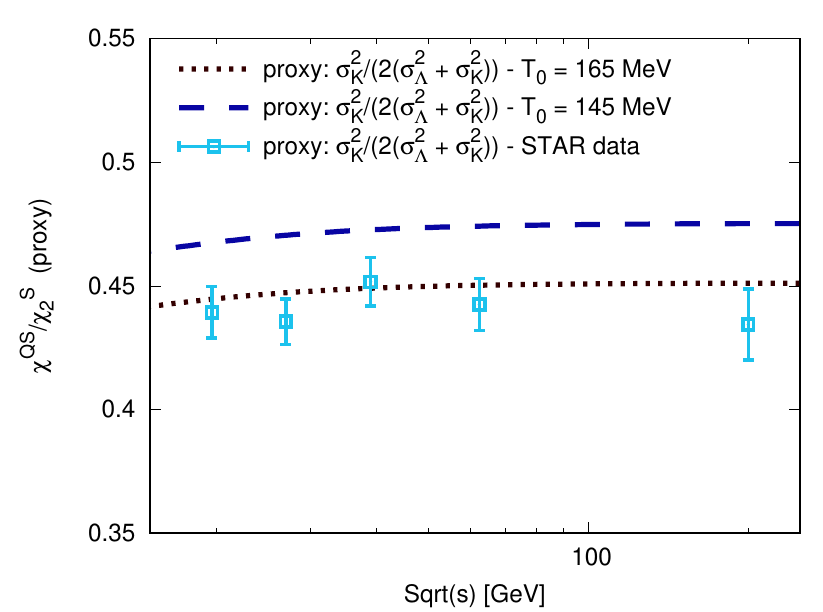}
\caption{Behavior of the proxies $\widetilde{C}^{\Lambda,\Lambda K}_{BS,SS}$ and $\widetilde{C}^{K,\Lambda K}_{QS,SS}$ along freeze-out lines with $T_0 = 145 \MeV$ (blue dashed line) and $T_0 = 165 \MeV$ (black dotted line) -- using different cuts for the different species, according to the experimental situation -- compared to the experimental results  \cite{Adamczyk:2017wsl,Nonaka:2019fhk} (light blue points).}
\label{fig:BS_SS_Proxy_Exp}
\end{figure}

In Fig. \ref{fig:BS_SS_Proxy_Exp} we show the behavior of the proxies
$\widetilde{C}^{\Lambda,\Lambda K}_{BS,SS}$ and $\widetilde{C}^{K,\Lambda
K}_{QS,SS}$ from Eqs. (\ref{eq:pxy_BS_SS_2}) and (\ref{eq:pxy_QS_SS_1}), along
the same freeze-out lines used in Fig. \ref{fig:BQS_ratios_pxy_FO}, and compare
them to available experimental results from the STAR Collaboration
\cite{Adamczyk:2017wsl,Nonaka:2019fhk}. The important difference is that now
the experimental cuts are the ones taken from the actual measurements, and
namely they are not the same for the different species. 

We see that the proxy (it is only one independent quantity as discussed above)
works well also in comparison with available experimental data, when the
considered freeze-out line is the one with a temperature $T ( \mu_B =0) = 165
\MeV$. This is in line with results from other analyses, which indicate that
strange particles seem to prefer a higher chemical freeze-out temperature
\cite{Bellwied:2018tkc}. 

One more remark is in order: by comparing, e.g. the curves in Figs.
\ref{fig:BS_SS_Proxy_Exp} (top panel) and \ref{fig:BQS_ratios_pxy_FO} (first
row, left or central panel), we can see how crucial it is that the same cuts
are applied to the different hadronic species utilized in a certain proxy. In
fact, the same ratio $\widetilde{C}^{\Lambda,\Lambda K}_{BS,SS}$ is shown, with
the difference that in Fig. \ref{fig:BQS_ratios_pxy_FO} the same cuts are
applied to both $\Lambda$ and $K$, while in Fig. \ref{fig:BS_SS_Proxy_Exp} the
cuts utilized are those from the experimental analyses, namely $0.9 < p_T < 2.0
\GeV, | y | < 0.5$ for net-$\Lambda$ \cite{Nonaka:2019fhk} and $0.4 < p_T < 1.6
\GeV, | y | < 0.5$ for net-kaon \cite{Adamczyk:2017wsl}. Due to this difference
in the applied kinematic cuts, more than a factor two separates the two curves.
For this reason, a direct comparison to lattice QCD would be premature.

\section{Conclusions} \label{sec:concl}

In this work, we first presented new continuum-extrapolated lattice QCD results
for second order non-diagonal correlators of conserved charges. While the
continuum extrapolation is a straightforward task at $\mu_B=0$, results need to
be extrapolated to the real $\mu_B$ regime, which cannot be simulated directly.
This is always ambiguous, so we compared two different schemes on a
for the ratio $\chi_{11}^{BS}/\chi_2^S$ and performed a
continuum extrapolation in the regime where the two approaches agree.

We performed an HRG-model-based study on the second order correlators, both diagonal and non-diagonal. At $\mu_B=0$ we found agreement with lattice.
Then we showed how they relate to fluctuations of those hadronic species which
can be measured in heavy ion collision experiments. What percentage of these
correlators is accounted for by particles that can actually be detected in the
experiment varies quite considerably from observable to observable.  

In order to compare either to lattice QCD results or experimental measurements, we focused on ratios of fluctuations, whose behavior can be reproduced through commonly measured hadronic observables, i.e. proxies. 

In the following we summarize the findings for a ratio with
each of the three possible cross-correlators of baryon ($B$), electric charge
($Q)$ and strangeness ($S$).

The $BQ$ correlator in equilibrium is dominated by proton fluctuations, with
the other contributions -- most notably the proton-pion correlation and
hyperons self correlations -- almost perfectly canceling each other.
Nonetheless, the information loss caused by isospin randomization prevents from
constructing successful proxies for ratios including $\chi_{11}^{BQ}$. 

Luckily, neither the isospin randomization, nor the introduction of cuts on the kinematics  had a significant effect on either $\chi_{11}^{BS}$ or $\chi_{11}^{QS}$. Because of this, we were able to construct proxies for the ratios $\chi_{11}^{BS}/\chi_2^S$ and $\chi_{11}^{QS}/\chi_2^S$ that are within $10\%$ of the grand canonical prediction. These two ratios are not independent, since in the isospin symmetric case they are related by the Gell--Mann-Nishijima formula.
It is striking that only two measured quantities, namely the variances of
net-kaon and net-Lambda distributions, were sufficient to build the
proxies for $\chi_{11}^{BS}/\chi_2^S$ and $\chi_{11}^{QS}/\chi_2^S$.
We showed that the inclusion of multi-strange hyperons does not improve the
quality of the proxy. Moreover, although these are cross correlators of
conserved charges, particle species cross correlators do not contribute
significantly. In fact, none of the particle cross correlators contributes to
any of the charge fluctuations or cross correlators, with the exception of the
proton-pion one, but the latter is largely affected by isospin randomization.

Thus, we have a ratio at hand that is available both from lattice simulations
and for experimental measurement. The ratio $\chi_{11}^{BS}/\chi_2^S$ behaves
as a strangeness-related thermometer for chemical freeze-out. 
We provided continuum extrapolated results at zero and finite chemical
potential for this quantity.

Finally, we compare our results to experiment.
A direct use of lattice data in the experimental context
would require the use of the same kinematic cuts for $\Lambda$ and $K$.
The STAR Collaboration has published results for fluctuations of $K$ and preliminary results for $\Lambda$ fluctuations, though with different kinematic cuts. To test our proxy, we recalculated its HRG model prediction with the actual cuts used in experiment. We saw that the $\sigma_\Lambda^2/( \sigma_K^2 + \sigma_\Lambda^2)$ ratio quite evidently favors the higher freeze-out temperature, in line with what was already shown by other analyses \cite{Bellwied:2018tkc,Bluhm:2018aei}. 

These high temperatures for the chemical freeze-out motivates the use of lattice QCD in future studies since they fall at the limit of the validity of the HRG model.

\section*{Acknowledgements}
This project was partly funded by the DFG grant SFB/TR55 and also
supported by the Hungarian National Research,  Development and
Innovation Office, NKFIH grants KKP126769 and K113034. The project
also received support from the BMBF grant 05P18PXFCA.
Parts of this work were supported  by  the National Science Foundation  under
grant  no.  PHY-1654219 and by the U.S.  Department of
Energy,  Office  of Science,  Office  of  Nuclear Physics, within the framework
of the Beam Energy Scan Theory (BEST) Topical Collaboration. 
A.P. is supported by the J\'anos Bolyai Research Scholarship of the
Hungarian Academy of Sciences and by the \'UNKP-19-4 New National Excellence
Program of the Ministry of Innovation and Technology.
The authors gratefully acknowledge the Gauss Centre for Supercomputing e.V.
(www.gauss-centre.eu) for funding this project by providing computing time on
the GCS Supercomputer JUWELS and JURECA/Booster at J\"ulich Supercomputing
Centre (JSC), and on SUPERMUC-NG at LRZ, Munich
 as well as on HAZELHEN at HLRS Stuttgart, Germany.  C.R. also
acknowledges the support from the Center of Advanced Computing and Data Systems
at the University of Houston. J.N.H. acknowledges the support of the Alfred P.
Sloan Foundation, support from the US-DOE Nuclear Science Grant No.
de-sc0019175. R.B. acknowledges support from the US DOE Nuclear Physics Grant
No. DE-FG02-07ER41521.



\appendix

\section{Tabulated lattice results}

In this appendix we give the results of the continuum extrapolations
explained in Fig.~\ref{fig:cont_ext_example} and plotted in Figs.~\ref{fig:BQ},
\ref{fig:QS} and \ref{fig:BS}.

\begin{table}[ht]
\begin{tabular}{|c|c|c|c|c|}
\hline
$T [\MeV] $& 
$\chi^{BQ}_{11}$ &
$\chi^{QS}_{11}$ &
$\chi^{BS}_{11}$ &
$\chi^{BS}_{11}/\chi^S_2$\\
\hline
130& 0.0086(11)& 0.0514(15)& -0.0132(14)& -0.1165(102)\\
135& 0.0101(08)& 0.0604(20)& -0.0167(17)& -0.1239(87)\\
140& 0.0124(08)& 0.0699(18)& -0.0227(13)& -0.1467(58)\\
145& 0.0162(12)& 0.0806(20)& -0.0332(18)& -0.1757(46)\\
150& 0.0217(17)& 0.0914(12)& -0.0491(28)& -0.2065(50)\\
155& 0.0242(10)& 0.1045(09)& -0.0676(38)& -0.2359(34)\\
160& 0.0266(07)& 0.1193(15)& -0.0825(27)& -0.2529(36)\\
165& 0.0278(06)& 0.1345(20)& -0.0981(26)& -0.2655(20)\\
170& 0.0277(04)& 0.1478(22)& -0.1136(23)& -0.2765(13)\\
175& 0.0269(04)& 0.1600(23)& -0.1296(24)& -0.2856(12)\\
180& 0.0256(05)& 0.1724(21)& -0.1455(23)& -0.2930(10)\\
185& 0.0242(04)& 0.1869(21)& -0.1610(22)& -0.2988(08)\\
190& 0.0227(04)& 0.2011(15)& -0.1749(24)& -0.3033(07)\\
\hline
\end{tabular}
\caption{\label{tab:latresult}
The continuum extrapolation of the cross-correlators
at $\mu_B=0$ from the lattice.
}
\end{table}

\bibliography{reference}

\end{document}